\documentclass[10pt, conference, letterpaper]{IEEEtran}
\IEEEoverridecommandlockouts

\usepackage[binary-units=true, group-separator={,}, group-minimum-digits={3}]{siunitx}
\usepackage{textcomp}
\usepackage{multirow}
\usepackage{graphicx}
\usepackage{subfig}
\usepackage{float}
\usepackage{url}
\usepackage{booktabs}
\usepackage{hyperref}
\usepackage{hyperxmp}
\usepackage{amsmath}
\usepackage{bbding}
\usepackage{gensymb}
\usepackage{pifont}
\usepackage{xspace}
\usepackage{tikz}
\usepackage{fp}
\usepackage{enumitem}
\usepackage[ruled,linesnumbered]{algorithm2e}
\usepackage{algpseudocode}
\usepackage{diagbox}
\usepackage{threeparttable}
\usepackage{caption}
\usepackage{ragged2e}
\usepackage[table,xcdraw]{xcolor}
\usepackage{cleveref}  
\usepackage{hhline}
\usepackage{ulem}
\crefformat{figure}{#2Figure~#1#3} 

\newlength\maxlen

\def\header{Acc.}
\newlength\maxlentime

\def\headertime{New}
\newlength\maxlenenergy

\def\headerenergy{New}
\newcommand{\System}{\textit{Stride}\xspace}
\newcommand{\Raw}{\textit{Raw}\xspace}

\begin{document}

\title{Revisiting-Aware In-Orbit Edge Computing \\ for Earth Observation}

\author{
\IEEEauthorblockN{
Zehua Sun$^\dag$,
Tao Ni$^\ddag$,
Kaiyan Cui$^\|$,
Weitao Xu$^{\S*}$,\thanks{* Weitao Xu is the corresponding author.}
Jingxian Wang$^\dag$
}
\IEEEauthorblockA{
$\dag$ National University of Singapore, Singapore
}
\IEEEauthorblockA{
$\ddag$ King Abdullah University of Science and Technology, Saudi Arabia
}
\IEEEauthorblockA{
$\|$ Nanjing University of Posts and Telecommunications, China
}
\IEEEauthorblockA{
$\S$ City University of Hong Kong, China
}
\vspace{-0.3in}
}

\maketitle
\pagestyle{plain}

\begin{abstract}
Typically, Earth observation satellites follow a rule of revisiting cycle to periodically pass over the same area of the Earth at regular intervals, which is jointly determined by their orbital properties (e.g., eccentricity, inclination) and instrument characteristics (e.g., off-nadir pointing and swath capabilities). However, we have observed delays in perceived revisiting cycles where limited satellite downlink bandwidth allows only partial images to be delivered, pushing back the timeliness of the full set of data, which we term as \textbf{revisiting cycle delay}. 

In this paper, we present a revisiting-aware in-orbit edge computing framework for Earth observation termed \System. \System leverages the unique orbital revisiting properties to afford historical reference revisiting images onboard, and exploits the inherent temporal redundancy in the revisiting imagery to transmit only the Regions of Interest (RoIs). Specifically, \System comprises a mono- and multi-temporal cloud indicator to alleviate cloud contamination, a coarse-to-fine reference selector for orbit deviation correction, and an ensemble-local change detector to address inter-band complexities and pixel-level perturbations. Experiments on a Flat-Sat testbed and a constellation simulator demonstrate \System improves the Revisiting Imagery Delivery (RID) score by up to 4.55$\times$, decreases the connectivity latency by 5.02$\times$, and enlarges the mapping coverage by 2.56$\times$, yielding state-of-the-art performance. 

\end{abstract}

\section{Introduction} \label{sec:Introduction}
The deployment of spacecraft in Earth's orbits has experienced a remarkable surge in recent years, resulting in an approximately $10\times$ increase in the number of satellites currently in orbit~\cite{web:Statista}, with well-known types as remote sensing (e.g., Landsat, Sentinel), communication (e.g., Starlink, OneWeb), and navigation (e.g., GPS, GLONASS). 
In remote sensing, Earth observation satellites are now equipped with state-of-the-art sensor arrays and sophisticated computer systems, empowering them to support a diverse range of applications (e.g., ground mapping, climate monitoring, and precision agriculture~\cite{van2021spacenet, toker2022dynamicearthnet}), and possess in-orbit computing capabilities for data processing in space~\cite{denby2020orbital, denby2023kodan}.

\begin{figure}[t]
    \centering
    \includegraphics[width=0.49\textwidth]{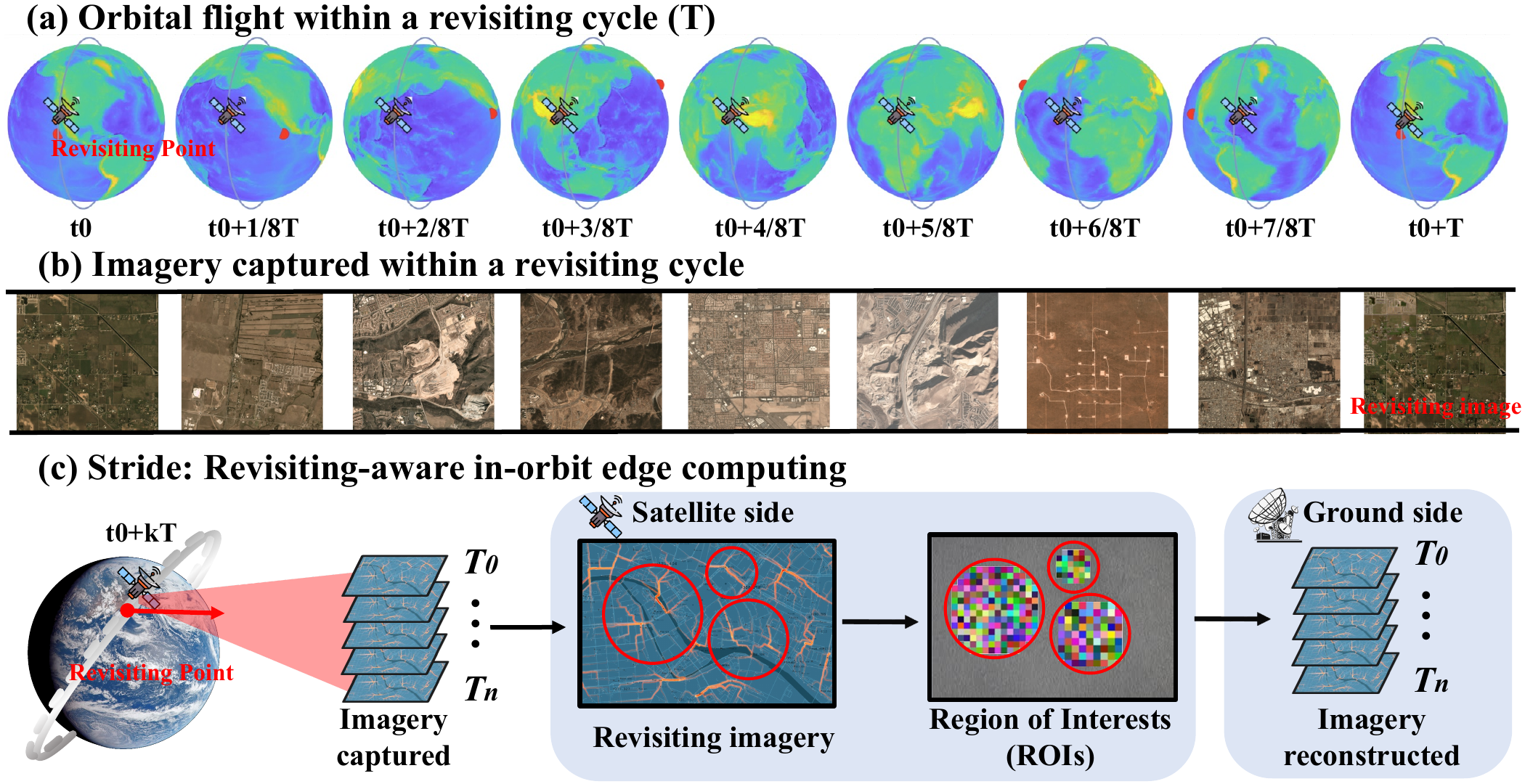}
    \vspace{-0.1in}
    \caption{
    \textbf{(a)} The revisiting cycle is a primary property of Earth observation satellites, whereby satellites repeatedly pass over the same regions of the Earth. 
    \textbf{(b)} As a satellite revisits a region, the new image often shares substantial temporal redundancy with prior observations.
    \textbf{(c)} \System exploits the inherent temporal redundancy among revisiting imagery to transmit only the RoIs, which significantly decreases connectivity latency (5.02$\times$) and enlarges mapping coverage (2.56$\times$). }
    \label{fig:System}
\end{figure}

Typically, Earth observation satellites follow a periodic path, repeatedly passing over specific regions to capture and transmit imagery to ground stations---a process termed revisiting cycle. 
Satellite revisiting cycles are jointly determined by orbital properties (e.g., eccentricity, inclination~\cite{russell1964kepler}) that dictate when the ground track recurs~\cite{nadoushan2015repeat}, and instrument characteristics (e.g., off-nadir pointing and swath capabilities) that can significantly outpace the orbital cycle by expanding the observable area per overpass.
For instance, NASA's Terra and Aqua satellites, equipped with Moderate Resolution Imaging Spectroradiometer (MODIS), achieve 1--2 day global revisit~\cite{justice2002overview}. 
However, we have observed a \textbf{revisiting cycle delay}, where limited satellite downlink bandwidth allows only partial images to be delivered, pushing back the timeliness of the full set of data. 
Specifically, despite the proliferation of observation satellites, it has been proved that the downlink capacity remains woefully inadequate to support massive satellite imagery data, rendering only 2\% of the captured data being received, due to high orbital dynamics and sparse ground station distributions~\cite{denby2020orbital, denby2023kodan}. 
The consequence is that, while a satellite can transmit the revisiting imagery from some specific nadir points to ground stations, it cannot transmit the full set of imagery of all nadir points along its track within the same revisiting cycle. 
This limitation results in a significant perceived revisiting cycle delay for some regions, potentially extending the revisiting cycle beyond the intended daily revisiting (e.g., from \SI{1}{} to \SI{1.5}{days}), which causes significant connectivity delays and limited mapping coverage in satellite observations. 
The impact is most pronounced in time-critical scenarios, such as wildfire outbreaks, sudden flooding, and impending hurricanes, where partial image delivery can severely compromise emergency response efforts.

\begin{table}[]
\centering
\caption{Comparison with the existing works.}
\renewcommand{\arraystretch}{1.1}
\resizebox{0.48\textwidth}{!}{%
\begin{tabular}{cllccccc}
\toprule[1.5pt]
\multicolumn{3}{c}{\textbf{Method}} &
  \textbf{\begin{tabular}[c]{@{}c@{}}Data \\      completeness\end{tabular}} &
  \textbf{\begin{tabular}[c]{@{}c@{}}Data \\      reliability\end{tabular}} &
  \textbf{\begin{tabular}[c]{@{}c@{}}Uplink \\      independence\end{tabular}} &
  \textbf{\begin{tabular}[c]{@{}c@{}}Orbital \\      properties\end{tabular}} &
  \textbf{\begin{tabular}[c]{@{}c@{}}Practical \\      deployability\end{tabular}} \\ \hline
\multicolumn{1}{c|}{} &
  \multicolumn{1}{l|}{} &
  \cellcolor[HTML]{EFEFEF}\textbf{OEC~\cite{denby2020orbital}} &
  \cellcolor[HTML]{EFEFEF}\ding{109} &
  \cellcolor[HTML]{EFEFEF}\ding{108} &
  \cellcolor[HTML]{EFEFEF}\ding{108} &
  \cellcolor[HTML]{EFEFEF}\ding{109} &
  \cellcolor[HTML]{EFEFEF}\ding{108} \\ \cline{3-8} 
\multicolumn{1}{c|}{} &
  \multicolumn{1}{l|}{} &
  \textbf{Kodan~\cite{denby2023kodan}} &
  \ding{109} &
  \ding{108} &
  \ding{108} &
  \ding{109} &
  \ding{108} \\ \cline{3-8} 
\multicolumn{1}{c|}{} &
  \multicolumn{1}{l|}{\multirow{-3}{*}{\textbf{\begin{tabular}[c]{@{}l@{}}Task-specific\\      filtering\end{tabular}}}} &
  \cellcolor[HTML]{EFEFEF}\textbf{Serval~\cite{tao2024known}} &
  \cellcolor[HTML]{EFEFEF}\ding{109} &
  \cellcolor[HTML]{EFEFEF}\ding{108} &
  \cellcolor[HTML]{EFEFEF}\ding{109} &
  \cellcolor[HTML]{EFEFEF}\ding{109} &
  \cellcolor[HTML]{EFEFEF}\ding{108} \\ \hhline{~=======}
\multicolumn{1}{c|}{} &
  \multicolumn{1}{l|}{} &
  \textbf{JPEG-2000~\cite{taubman2002jpeg2000}} &
  \ding{108} &
  \ding{108} &
  \ding{108} &
  \ding{109} &
  \ding{108} \\ \cline{3-8} 
\multicolumn{1}{c|}{} &
  \multicolumn{1}{l|}{\multirow{-2}{*}{\textbf{\begin{tabular}[c]{@{}l@{}}Conventional\\      compression\end{tabular}}}} &
  \cellcolor[HTML]{EFEFEF}\textbf{CCSDS~\cite{CCSDS121B3}} &
  \cellcolor[HTML]{EFEFEF}\ding{108} &
  \cellcolor[HTML]{EFEFEF}\ding{108} &
  \cellcolor[HTML]{EFEFEF}\ding{108} &
  \cellcolor[HTML]{EFEFEF}\ding{109} &
  \cellcolor[HTML]{EFEFEF}\ding{108} \\ \hhline{~=======}
\multicolumn{1}{c|}{} &
  \multicolumn{1}{l|}{} &
  \textbf{RTCS~\cite{hsu2024real}} &
  \ding{108} &
  \ding{109} &
  \ding{108} &
  \ding{109} &
  \ding{109} \\ \cline{3-8} 
\multicolumn{1}{c|}{\multirow{-7}{*}{\textbf{\rotatebox{90}{Mono-level}}}} &
  \multicolumn{1}{l|}{\multirow{-2}{*}{\textbf{\begin{tabular}[c]{@{}l@{}}Deep   learning \\      compression\end{tabular}}}} &
  \cellcolor[HTML]{EFEFEF}\textbf{DeepSpace~\cite{sun2025deepspace}} &
  \cellcolor[HTML]{EFEFEF}\ding{108} &
  \cellcolor[HTML]{EFEFEF}\ding{109} &
  \cellcolor[HTML]{EFEFEF}\ding{108} &
  \cellcolor[HTML]{EFEFEF}\ding{109} &
  \cellcolor[HTML]{EFEFEF}\ding{109} \\ \hhline{========}
\multicolumn{1}{c|}{} &
  \multicolumn{1}{l|}{\textbf{\begin{tabular}[c]{@{}l@{}}Reference UL \&\\      compression DL\end{tabular}}} &
  \textbf{Earth+~\cite{du2025earth+}} &
  \ding{109} &
  \ding{108} &
  \ding{109} &
  \ding{109} &
  \ding{109} \\ \hhline{~=======}
\multicolumn{1}{l|}{\multirow{-2}{*}{\rotatebox{90}{\makebox[0.4cm][c]{ \textbf{Multi-level}}}}}  &
  \multicolumn{1}{l|}{\textbf{\begin{tabular}[c]{@{}l@{}}Orbital   revisiting\\      properties\end{tabular}}} &
  \cellcolor[HTML]{EFEFEF}{\color[HTML]{000000} \textbf{\textbf{\System (ours)}}} &
  \cellcolor[HTML]{EFEFEF}{\color[HTML]{000000} \ding{108}} &
  \cellcolor[HTML]{EFEFEF}{\color[HTML]{000000} \ding{108}} &
  \cellcolor[HTML]{EFEFEF}{\color[HTML]{000000} \ding{108}} &
  \cellcolor[HTML]{EFEFEF}{\color[HTML]{000000} \ding{108}} &
  \cellcolor[HTML]{EFEFEF}{\color[HTML]{000000} \ding{108}} \\ 
\bottomrule[1.5pt]
\end{tabular}%
}
\begin{tablenotes}
\item \textit{(\ding{108}--satisfactory, \ding{109}--weak).}
\end{tablenotes}
\vspace{-0.2in}
\label{tab:Comparison}
\end{table}

As illustrated in~\autoref{tab:Comparison}, despite extensive efforts to alleviate data transmission bottlenecks, the existing solutions remain limited in data reliability, practical deployment, etc. 
Specifically, the existing solutions typically operate at the mono-imagery level (i.e., task-specific filtering~\cite{denby2020orbital, denby2023kodan}, and mono-imagery compression~\cite{taubman2002jpeg2000, CCSDS121B3, hsu2024real, sun2025deepspace}) and the multi-imagery level (i.e., reference uplinking-based compression~\cite{du2025earth+}). 
\textbf{(1) Task-specific filtering.} The emerging satellite orbital edge computing systems (e.g., OEC~\cite{denby2020orbital}, Kodan~\cite{denby2023kodan}, and Serval~\cite{tao2024known}) perform task-specific and high-value data filtering or queries, but sacrifice data completeness by preserving only selected information. 
\textbf{(2) Conventional compression.} 
The conventional mono-imagery compression methods (e.g., JPEG-2000~\cite{taubman2002jpeg2000}, CCSDS 121.0-B-3~\cite{CCSDS121B3}) predominantly exploit spatial or spectral redundancy within individual images, but fail to exploit the substantial temporal redundancy inherent in satellite imagery, resulting in limited compression performance. 
\textbf{(3) Deep learning compression.} 
The recent deep learning-based ones (e.g., RTCS~\cite{hsu2024real}, DeepSpace~\cite{sun2025deepspace}) typically train neural networks to learn a compact latent representation of imagery for compression, which is then deployed on satellites for encoding, and on the ground for reconstruction. 
However, they exhibit poor generalization, often failing to handle unseen object categories or novel scenes, rendering unreliable data for operators. 
\textbf{(4) Reference uplink \& compression downlink.} 
A recent advancement is Earth+~\cite{du2025earth+}, where the ground stations pick and then uplink the reference imagery to satellites, while the satellites use the references to compress and then downlink the compressed blocks in the images to the ground. 
Although effective in exploiting multi-temporal redundancy, it requires frequent reference selection and uplinks for onboard comparison, rendering it impractical for satellite systems with highly asymmetric uplink and downlink capacities.

As the first attempt from the perspective of celestial mechanics, we present a revisiting-aware in-orbit edge computing framework for Earth observation termed \System. 
The key rationale of \System is to leverage the unique orbital revisiting properties of the satellite to afford historical reference revisiting images, and exploit the inherent temporal redundancy in the revisiting imagery to transmit only the Regions of Interest (RoIs). 
However, to design and implement \System, there are three main challenges: cloud cover contamination, orbit deviation, and inter-band complexities \& pixel perturbations. 
To resolve these challenges, we further propose the mono- and multi-temporal cloud indicator, coarse-to-fine reference selector, and ensemble-local change detector. 
As illustrated in~\autoref{fig:System}, for each captured image, \System first utilizes the cloud indicator to tackle cloud cover contamination, then employs the reference selector to frame the reference revisiting image, with which \System can identify RoIs of the revisiting imagery based on the change detector.




To design and implement \System, we have overcome the following three main challenges:

\noindent
\textbf{Challenge 1: Cloud cover contamination.}
Earth observation satellite imagery frequently suffers from cloud cover contamination~\cite{li2022cloud}, which severely degrades imagery quality. 
Although cloud detection methods are well studied, including spectral feature methods~\cite{zhu2015improvement, wang2006cloud} and deep learning-based ones~\cite{jeppesen2019cloud, shao2019cloud}, these methods typically remain confined to independent analysis of single or multi-temporal imagery over fixed regions. 
However, they overlook the richer perspective of satellite orbit dynamics, that is, they rarely adopt the satellite’s own orbital perspective, relying instead on ground-based viewpoints. 
In contrast, both inherent spatial geographical continuity along the orbit, and their associated temporal continuity can be jointly exploited (see~\autoref{sec:Cloud Indicator}). 
To this end, we propose a novel mono- and multi-temporal cloud indicator that explicitly leverages the inherent \textbf{spatial-temporal geographical continuity} of captured imagery along the satellite orbit.
\System initially processes the mono-temporal candidate image by exploiting the distinctive color and brightness characteristics of cloud cover in specific spectral bands for cloud detection. 
To further mitigate interference from other highly reflective objects (e.g., snowy mountains, ice sheets), \System leverages two complementary principles: 
(1) \textbf{Spatial continuity}: snow and ice exhibit widespread, uniform aggregation (e.g., polar regions), unlike scattered clouds—\System maintains a cloud cache tracking $n$ recent labels for each location; 
(2) \textbf{Temporal continuity}: when the cache fills with consecutive ``cloudy'' labels, \System triggers multi-temporal verification by comparing the candidate image against a temporally-distant reference—persistent high similarity indicates stationary terrain (false positive), while low similarity confirms transient cloud cover (true positive). 

\noindent
\textbf{Challenge 2: Orbit deviation.}
Deviations in satellite orbits inevitably occur (as elaborated in~\autoref{fig:Revisiting_Discrepancy}), caused by various factors such as orbital decay, Earth's gravitational perturbations, and solar radiation pressure~\cite{weinberg1986orbital}. 
The common solutions are orbit control (e.g., station-keeping using low-thrust maneuvers) and attitude control (e.g., hardware compensation and calibration) systems. 
However, long-duration invariance remains unattainable due to the intermittent controls and the saturation of momentum wheels (i.e., temporally limiting control)~\cite{weiss2018station}. 
Thus, we propose a coarse-to-fine reference selector to achieve strict satellite imagery alignment at the software level. 
\System first conducts a coarse-level query to identify potential reference image sets based on embedded geographical coordinate labels. 
However, fine alignment poses further challenges, primarily stemming from nonlinear changes (e.g., time-induced alterations, scale distortions) in imagery content over time. 
The existing methods either fall short of efficient computation or struggle with matching distortions (see~\autoref{sec:Fine Alignment}). 
To address these problems, we propose a \textbf{local-global consensus} method. 
Specifically, we first derive a \textbf{local geometric prior} by extracting the local feature points from the image pairs for keypoint matching. 
Subsequently, we establish the \textbf{global consensus} by leveraging the relative positions of these matching keypoint groups in their respective images for voting to achieve fine alignment. 

\noindent
\textbf{Challenge 3: Inter-band complexities \& pixel perturbations.}
With the reference revisiting images, change detection is performed to identify RoIs. 
Although the existing pixel-level change detection methods~\cite{cheng2024change} can capture temporal variations, they face two key challenges: (1) the multi-channel characteristics of satellite imagery introduce complex inter-band correlations, and (2) pixel perturbations can still arise due to optical hardware imperfections and environmental variations (e.g., illumination and atmospheric disturbances), which unnecessarily inflate the range of RoIs. 
To address these challenges, we propose an ensemble-local change detector. 
Specifically, \System first performs pixel-level ensemble differencing within a single channel to identify RoIs with temporal discrepancies, capitalizing on the \textbf{``one-to-many'' notion} that pixel-point variations across layers are correlated. 
We subsequently design \textbf{local compensation techniques} to refine the RoIs by distinguishing them from Regions of Noise (RoNs) and sparse representation.



\begin{figure}
\begin{minipage}{0.545\linewidth}
        \centering
        \includegraphics[width=\textwidth]{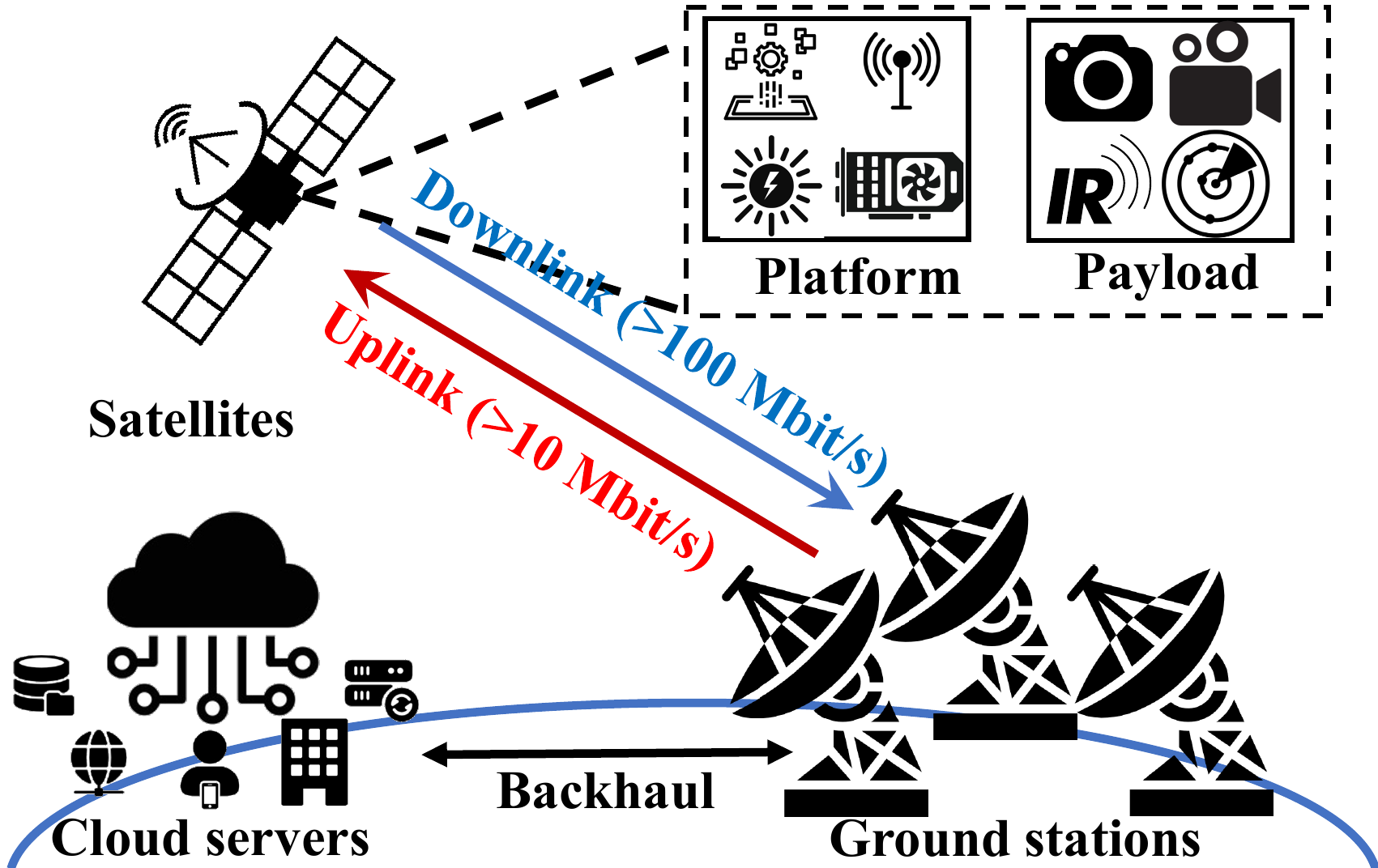}
        \vspace{-0.17in}
        \caption{Satellite systems.}
        \vspace{-0.05in}
        \label{fig:Satellite}
\end{minipage}
\hfill
\begin{minipage}{0.445\linewidth}
        \centering
        \includegraphics[width=\textwidth]{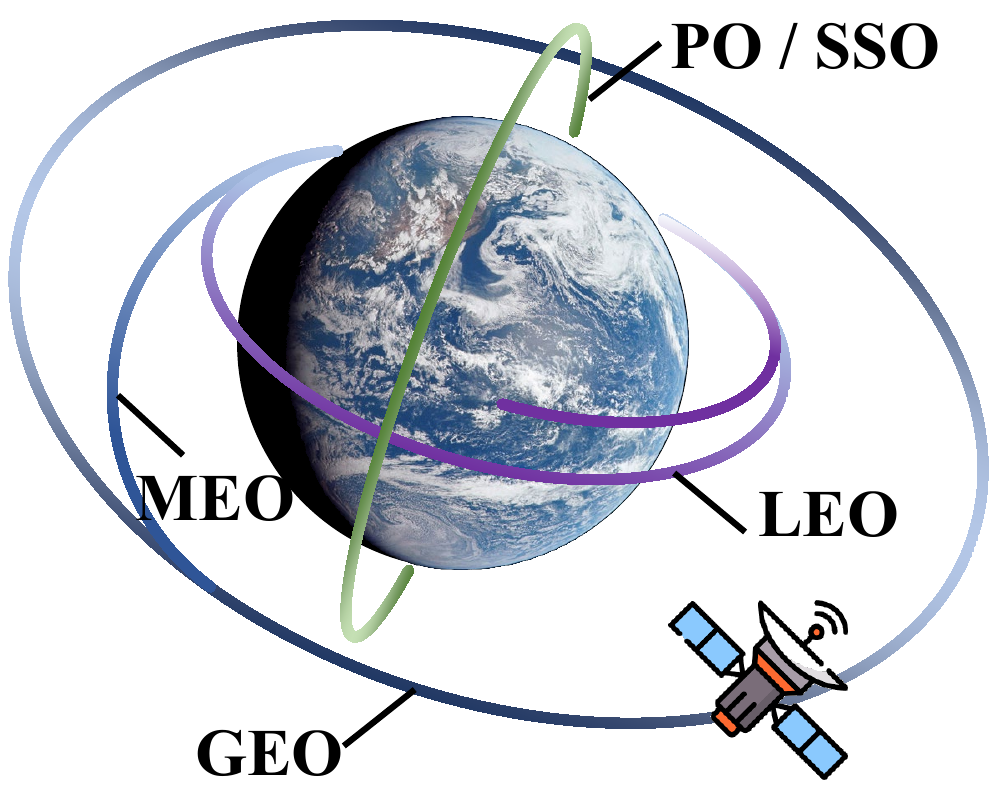}
        \vspace{-0.2in}
        \caption{Space orbits.}
         \vspace{-0.05in}
        \label{fig:Orbit}
\end{minipage}
\end{figure}

We implement and evaluate \System on a real-world Flat-Sat testbed and a constellation simulator.  
The Flat-Sat testbed integrates key flight subsystems in real satellites, where NVIDIA Jetson TX2 serves as the on-board computing (OBC) platform. 
The constellation simulator models orbital dynamics, data acquisition, and satellite communication behaviors. 
We evaluate Stride across five popular applications with three well-known observation satellite Two-Line Element (TLE) orbit descriptors. 
Experimental results reveal that: 
\begin{itemize}[leftmargin=*, itemindent=0em]
\item \System improves the Revisiting Imagery Delivery (RID) score (the metric to measure the revisiting capability as illustrated in~\autoref{sec:Revisiting Cycle Delay Phenomenon}) by up to 4.55$\times$ while maintaining the imagery reconstruction quality. 

\item \System decreases the connectivity latency by 5.02$\times$ and enlarges the mapping coverage by 2.56$\times$. 

\item \System yields state-of-the-art performance. 
\end{itemize}

\noindent
\textbf{Contributions.} \System makes the following contributions: 
\begin{itemize}[leftmargin=*, itemindent=0em]

\item To the best of our knowledge, \System is the first framework from the perspective of in-orbit revisiting properties in celestial mechanics. 

\item \System integrates a mono- and multi-temporal cloud indicator, a coarse-to-fine reference selector, and an ensemble-local change detector for various challenges. 





\end{itemize}


\section{Preliminary} \label{sec:Preliminary}
\noindent
\textbf{Earth observation satellite system.}
As shown in~\autoref{fig:Satellite}, an Earth observation satellite system comprises satellites, ground stations, and the cloud. 
Satellites capture imagery and transmit it to ground stations via downlink, which then relay data to cloud platforms (e.g., Google Earth Engine, ArcGIS Online) for storage and application services.

\noindent
\textbf{Uplink/downlink.}
Earth observation satellite transmission has an extremely asymmetric uplink and downlink. 
Specifically, the link frequency bands are primarily concentrated in the 7/8 GHz X band~\cite{denby2023kodan}.
The narrowband uplink typically reaches tens to hundreds of kbps for telecommand, while the wideband downlink can achieve hundreds of Mbps to tens of Gbps~\cite{tao2023transmitting} for payload data transmission.



\noindent
\textbf{Satellite orbit.}
As shown in~\autoref{fig:Orbit}, satellite orbits are categorized by altitude into three types: Low Earth Orbit (LEO), Medium Earth Orbit (MEO), and Geostationary Orbit (GEO). 
Additionally, two special orbits in LEO are critical for Earth observation: Polar Orbits (POs) and Sun-Synchronous Orbits (SSOs)~\cite{treblow2024responsive}. 
POs pass over the poles to provide global coverage.
SSOs are near-polar orbits, enabling satellites to revisit the same location at the same local solar time. 

\begin{figure}[t]
\vspace{-0.1in}
\centering
    \subfloat[RAAN, AOP, and inclination. ]{\includegraphics[width=0.24\textwidth]{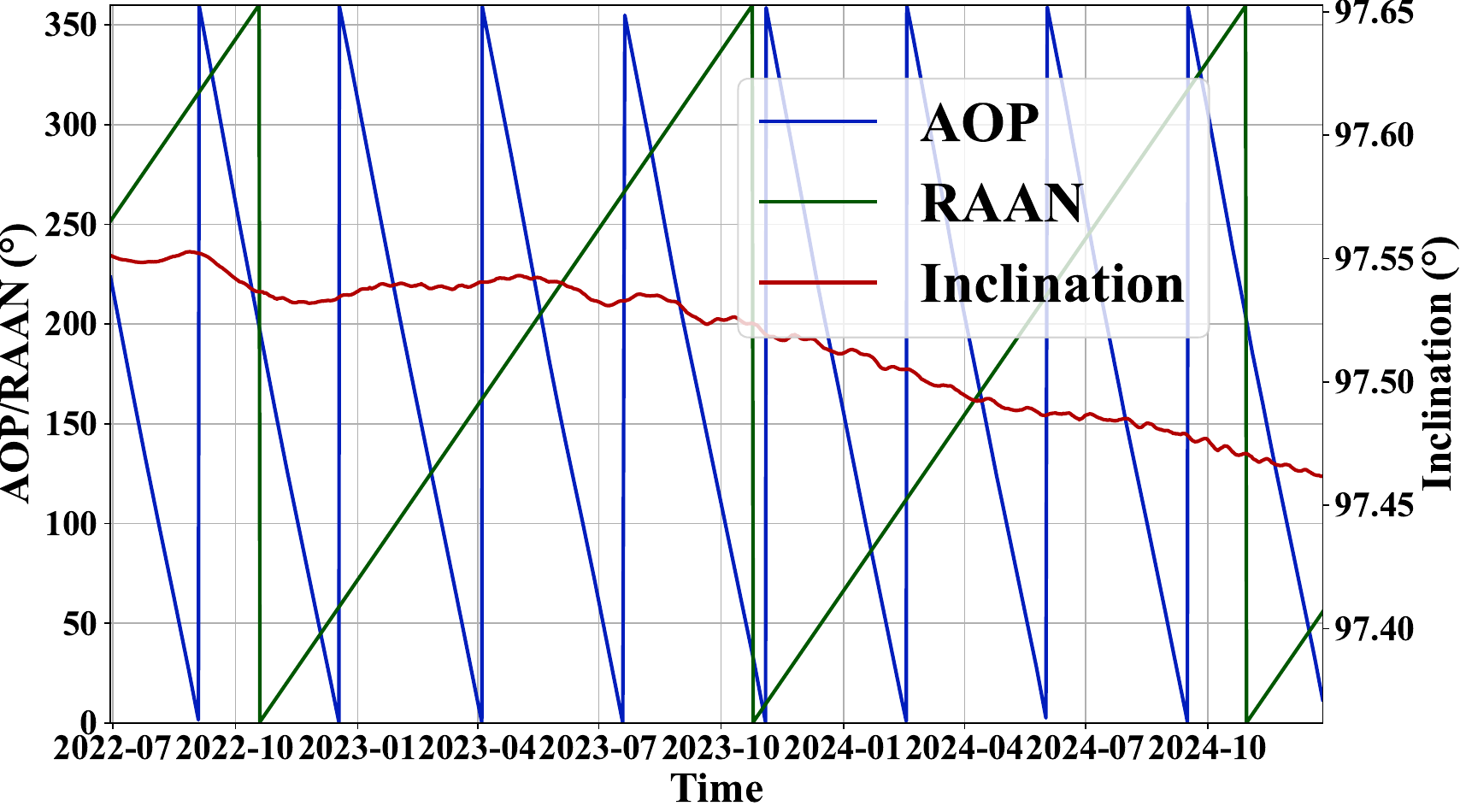}  \label{fig:Revisiting_Discrepancy1}}
     \subfloat[SMA and eccentricity.]{\includegraphics[width=0.24\textwidth]{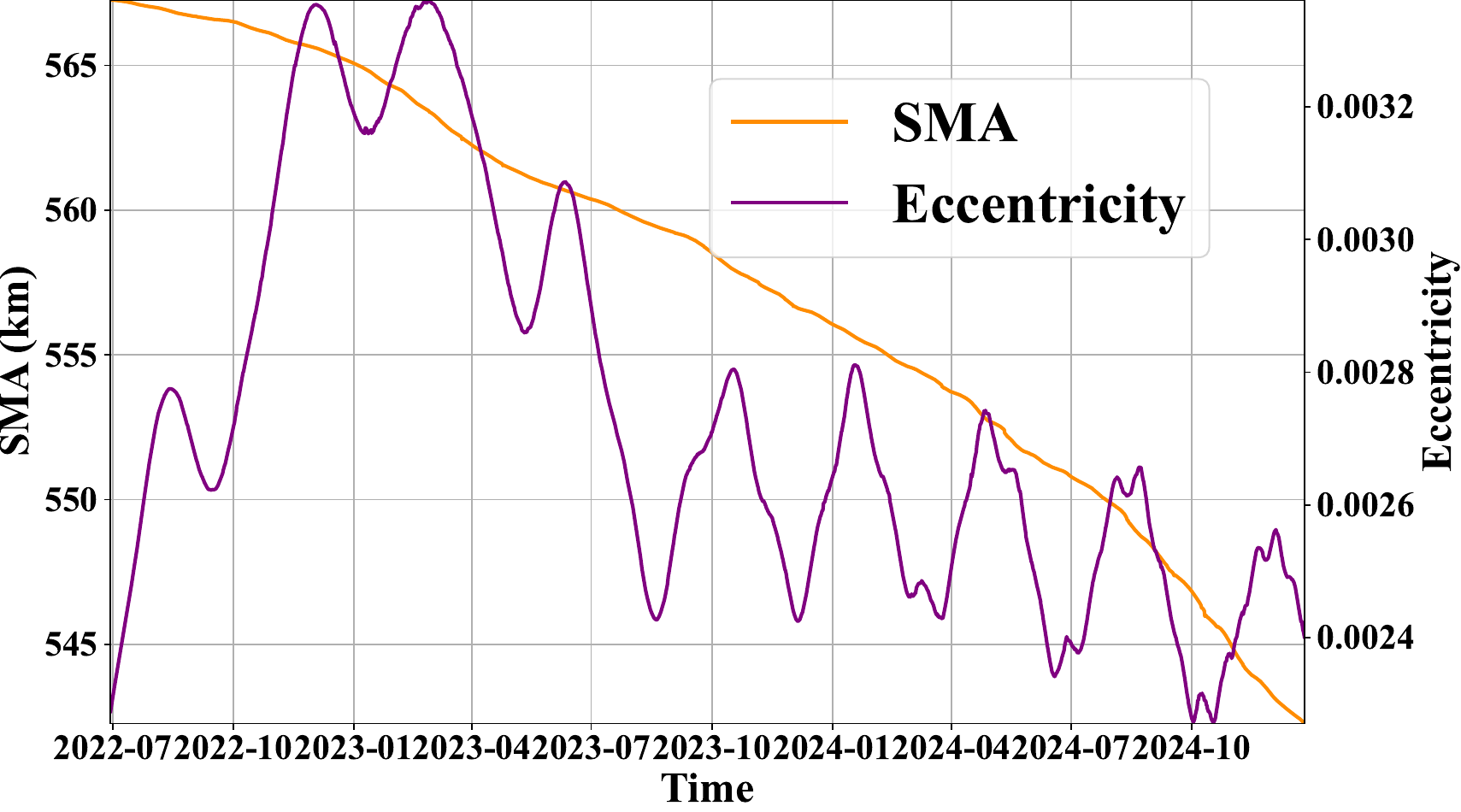}  \label{fig:Revisiting_Discrepancy2}}
 \vspace{-0.05in}
\caption{Illustration of the orbital elements variation over time, indicating the presence of orbit deviations.}
\vspace{-0.1in}
\label{fig:Revisiting_Discrepancy}
\end{figure}

\noindent
\textbf{Revisiting property.}
The revisiting property of satellites is jointly determined by their orbital properties and instrument characteristics~\cite{russell1964kepler}. 
As illustrated in~\autoref{fig:Revisiting_Discrepancy}, the orbital properties determine the orbital repeat cycle, dictating when the satellite's ground track recurs. 
Specifically, orbital elements consist of six parameters, including Semi-Major Axis (SMA) $a$, eccentricity $e$, inclination $i$, Argument of Perigee (AOP) $\omega$, Right Ascension of the Ascending Node (RAAN) $\Omega$, and mean anomaly $M$. 
These elements uniquely define a satellite's Keplerian orbit, where $a$, $e$, and $i$ determine its size, shape, and orientation, while $\omega$, $\Omega$, and $M$ specify its position along the orbit. 
It is noted that the orbital elements are not stable and exhibit long-term fluctuations, leading to a progressive drift in the satellite's orbit. 
On the basis, instrument characteristics, such as swath width and off-nadir pointing capability, expand the observable area per pass, enabling coverage from adjacent orbits and thus reducing revisit time beyond orbital constraints.

\section{\large Revisiting Cycle Delay Phenomenon} \label{sec:Revisiting Cycle Delay Phenomenon}
We quantify the scale of the revisiting cycle delay phenomenon. 
Due to the varying revisiting cycles across different satellites, we define a unified metric, the RID score, to enable fair comparison. 
Specifically, the RID score quantifies revisiting capability over a fixed time window (e.g., 24 hours) as the ratio of successfully delivered data to the total collected data along the satellite’s ground tracks. 

\begin{figure}[t]
\vspace{-0.15in}
\centering
	\subfloat[RID score over ground stations. ]{\includegraphics[width=0.24\textwidth]{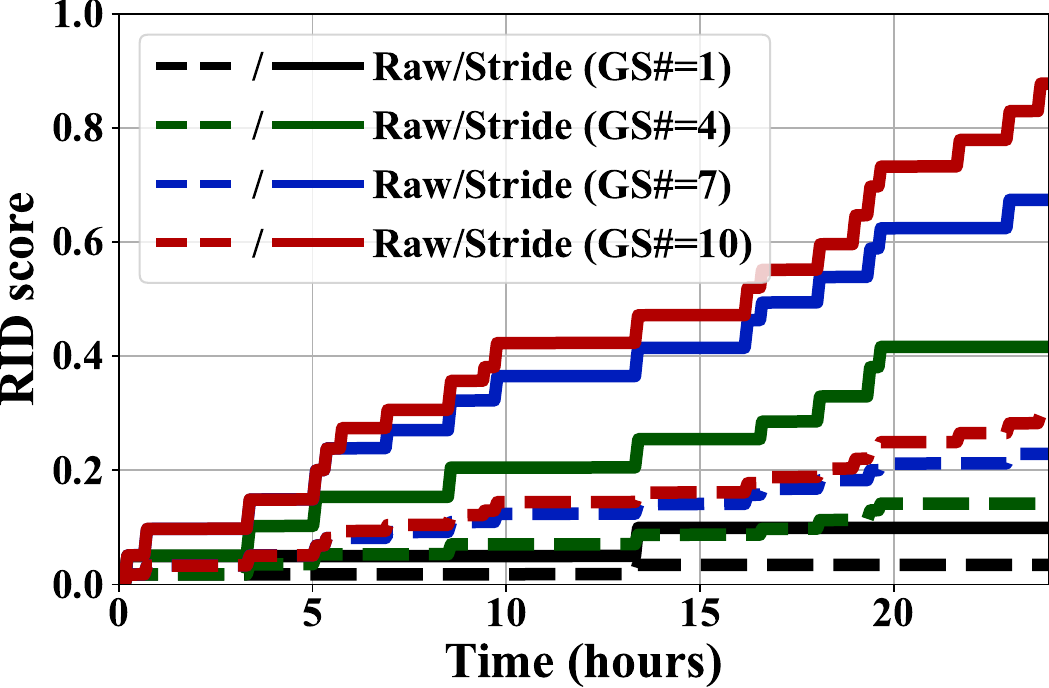} \label{fig:Partial_RID1}}
	\subfloat[Downlink distribution.]{\includegraphics[width=0.24\textwidth]{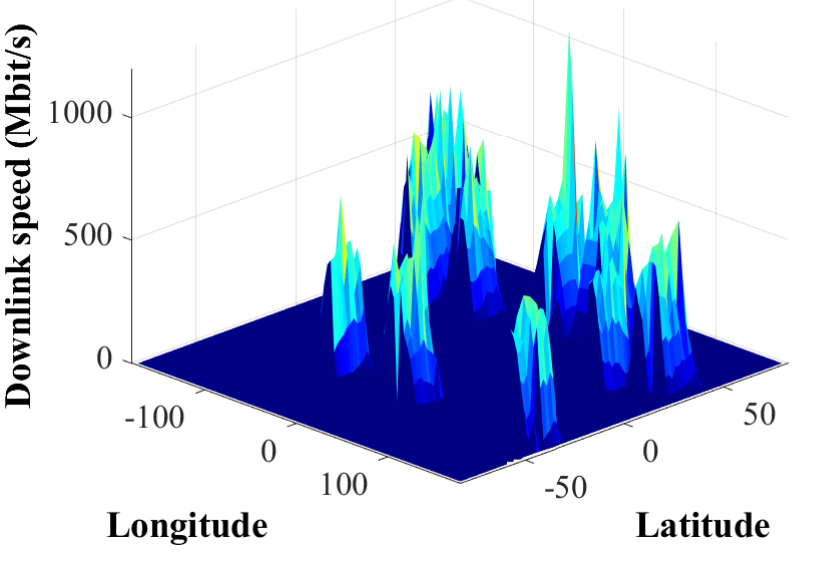} \label{fig:Partial_RID2}}
\vspace{-0.05in}
\caption{Quantification of revisiting cycle delay.}
\vspace{-0.1in}
\label{fig:Partial_RID}
\end{figure}

\cref{fig:Partial_RID1} illustrates the RID score over the number of ground stations. 
It is evident that the volume of data transmitted by the satellite is significantly lower than the total volume of data collected. 
With ten ground stations, the RID score is only 0.2982 (in contrast to 0.8776 for \System), which means that the revisiting cycles of 29.82\% of the regions are not delayed. 
This phenomenon is further exacerbated by the restricted number of ground stations, as depicted in~\cref{fig:Partial_RID2}.

\section{\System Design} \label{sec:System Design}
\subsection{Overview} \label{sec:Overview}
\begin{figure*}[t]
    \centering
    \includegraphics[width=0.98\textwidth]{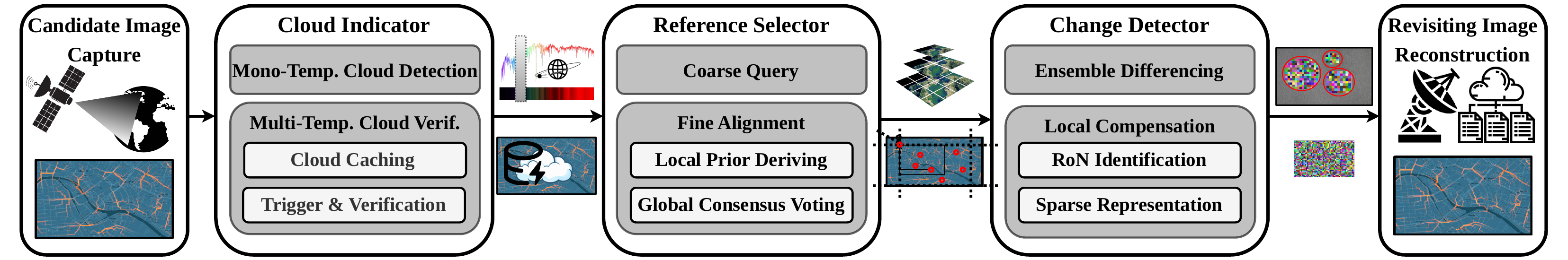}
    \vspace{-0.05in}
    \caption{Overview of \System (\autoref{sec:Overview}), consisting of a cloud indicator, a reference selector, and a change detector. }
    \vspace{-0.1in}
    \label{fig:System_Overview}
\end{figure*}

\autoref{fig:System_Overview} presents an overview of \System. 
When the satellite captures a revisiting image, it is processed by \System to identify RoIs for transmission, which are then reconstructed on the ground station side. 
Specifically, \System consists of three main components: the cloud indicator (\autoref{sec:Cloud Indicator}), the reference selector (\autoref{sec:Reference Selector}), and the change detector (\autoref{sec:Change Detector}).

\subsection{Cloud Indicator} \label{sec:Cloud Indicator}

\subsubsection{Mono-Temporal Cloud Detection} \label{sec:Mono-Temporal Cloud Detection} 
\noindent
When a satellite captures an image, it is typically stored in Tagged Image File Format (TIFF). 
In matrix representation, each element signifies a pixel value within the image. 
Thus, its three-dimensional matrix is denoted as $\mathcal{I}^{(H \times W \times N)}$, where $H$, $W$, and $N$ represent the height, width, and channel number of the image, respectively. 
The element $\mathcal{I}_{(h,w,n)}$ represents the pixel value at a specific height $h$, width $w$, and channel index $n$. 



For the captured revisiting image, we perform cloud detection on this mono-temporal candidate image $\mathcal{I}$ to determine its cloud coverage score. 
It is noted that cloud shadows are not considered in this process, because the texture features of objects obscured by the cloud shadow remain largely intact. 
Clouds typically render high reflectivity in some specific spectral bands~\cite{li2022cloud}. 
For the most common visible spectrum, clouds generally appear as bright features in the image due to the high reflectivity. 
Therefore, we adopt a threshold-based cloud detection method to capitalize on this characteristic, where cloud pixels exhibit high values in the channels of Red, Green, and Blue. 
We classify the pixel as part of a cloud if the values of all three channels exceed this threshold ratio $\tau_{mono}$. 
The cloud masking process can be represented as: 
\begin{equation}
\small
\mathcal{M}_{(h,w)}= \begin{cases} 1, & \text{if } \mathcal{I}_{(h,w,0:2)}>max(\mathcal{I})\times \tau_{mono} \\ 
0, & \text{otherwise. } \end{cases} 
\end{equation}
By comparing pixel values against the threshold, we can quantify the cloud coverage score by generating a binary cloud mask, which indicates the spatial extent of cloud formations: $S_{cloud} \leftarrow \sum{M}/{|M|}$. 
The image with a cloud coverage score exceeding 0.25, is classified as ``\textit{cloudy}'' and transmitted directly to ground stations.



\subsubsection{Multi-Temporal Cloud Verification} \label{sec:Multi-Temporal Cloud Verification} 

\noindent
\textbf{Cloud cache.}
As mentioned previously, other highly reflective surfaces, particularly in snowy landscapes and ice sheets, may influence the cloud coverage score. 
To address this issue, we propose a multi-temporal cloud verification method as a supplement. 
We leverage the distinctive spatial continuity that expansive white snow and ice terrain typically exhibits extensive aggregation (e.g., in the Arctic and Antarctica regions), in contrast to the scattered nature of cloud cover. 
Subsequently, we devise a cache to store each cloud label of the captured revisiting image over time. 
Essentially, if the cache consistently records multiple ``\textit{cloudy}'' labels, it signifies the potential presence of a prominent snowy or icy landscape rather than the cloud cover. 


\noindent
\textbf{Trigger \& verification.}
The cloud cache records the cloud labels of the $n$ most recent revisiting images. 
When the cloud cache is filled with ``\textit{cloudy}'' labels, we perform multi-temporal cloud verification by considering the temporal continuity. 
Specifically, for such a candidate image, rather than direct transmission, it is processed by the reference selector (\autoref{sec:Reference Selector}) and then the change detector (\autoref{sec:Change Detector}). 
The selected reference is the image of the same revisiting area but captured at different times with the candidate image. 
Thus, the similarity between these two multi-temporal images can be utilized to validate the label of ``\textit{cloudy}'', leveraging the dynamic nature of cloud formations. 
We indirectly represent this similarity $\tau_{multi}$ through the differencing ratio in the change detector. 
In cases where the candidate image labeled as ``\textit{cloudy}'' exhibits a relatively high similarity score, it suggests that other highly reflective objects are misidentified as clouds. 
Conversely, for an actual ``\textit{cloudy}'' image, due to the drowning of useful features in the image, the reference will be selected incorrectly, or the similarity will be relatively low. 
This multi-temporal cloud verification aids in distinguishing between actual cloud cover and other objects present in the image, which ensures a more reliable cloud assessment in satellite imagery analysis.

\subsection{Reference Selector} \label{sec:Reference Selector}

\subsubsection{Coarse Query} \label{sec:Coarse Query} 
\noindent
Satellite images are typically embedded with geographical information of latitude and longitude coordinates, which allows for a coarse query to filter potential images by imposing restrictions on these geographic labels.
In this context, given a candidate image $\mathcal{I}$ with its nadir point coordinates $\left(\lambda_{\mathcal{I}}, \phi_{\mathcal{I}}\right)$, we identify potential reference image groups by constraining its latitude and longitude coordinates within a predefined range $\left(\tau_\lambda, \tau_\phi\right)$. 
Among them, adjacent potential reference images are concatenated to form larger reference images for fine alignment.

\subsubsection{Fine Alignment} \label{sec:Fine Alignment} 
\noindent
\textbf{Local prior deriving.}
We utilize Speeded Up Robust Features (SURF)~\cite{bay2006surf} to detect points of interest (referred to as keypoints), by constructing a Hessian matrix~\cite{thacker1989role}, which typically identifies image edges or regions with rapid intensity changes. 
For the given candidate image frame $\mathrm{I}$, we utilize its first channel for consideration of energy efficiency. 
We then construct its Hessian matrix $\mathbf{H}$ after applying Gaussian filters, which is represented as:
\begin{equation}
\small
\mathbf{H}(w, \sigma)=\left(\begin{array}{ll}
L_{w w}(w, \sigma) & L_{w h}(w, \sigma) \\
L_{h w}(w, \sigma) & L_{h h}(w, \sigma)
\end{array}\right),
\end{equation}
where $L_{w w}, L_{w h}, L_{h w}$, and $L_{h h}$ are Gaussian second-order derivatives at point $w$ and scale $\sigma$.


SURF approximates the Hessian matrix using box filters $D$, allowing for efficient computation through integral images:
\begin{equation}
\small
\operatorname{Det}\left(\mathbf{H}_{\text {approx}}\right)=D_{w w} D_{h h}-\left(\alpha D_{w h}\right)^2,
\end{equation}
where $D_{w w}, D_{h h}$, and $D_{w h}$ are approximations using box filter, and $\alpha$ is a weighting factor to balance the filter responses.

Subsequently, SURF creates a scale space by upscaling the filter size rather than downscaling the image to initially generate keypoints. 
SURF assigns a dominant orientation to each keypoint based on the Haar wavelet~\cite{stankovic2003haar} responses within a circular region surrounding the keypoint.

After keypoints detected and oriented, SURF constructs a feature descriptor for each keypoint by analyzing the intensity changes in its local neighborhood, specifically using Haar wavelet responses. 
The local neighborhood is divided into several square sub-regions. 
For each sub-region, SURF calculates the Haar wavelet responses in both horizontal ($d_w$) and vertical ($d_h$) directions. 
The descriptor for each sub-region consists of the sums of these responses:
\begin{equation}
\small
\mathbf{d} = \left( \sum d_w, \sum d_h, \sum |d_w|, \sum |d_h| \right).
\end{equation}

Keypoints in the candidate image and the reference image are considered matches if their feature descriptors are sufficiently similar, measured by the Euclidean distance.

\noindent
\textbf{Global consensus voting.}
After keypoints matching, we perform fine alignment steps.  
The SURF matching method relies on the homography of the matched keypoints, which is a perspective transformation between two planes. 
However, this transformation can introduce distortions or deformations in the matching region, significantly affecting the accuracy of the change detector. 
To address this issue, we propose a global consensus fine alignment method via origin voting. 
Leveraging the fact that the revisiting reference image shares the same size as the candidate image, we compute the origin index of the matching region in the reference image (i.e., the top-left corner). 
For each matched keypoint group $k (w,h)$ and $k^\prime(w^\prime, h^\prime)$, we calculate the origin index of the matching region in the reference image based on the relative location of its corresponding keypoint in the candidate image via $k^\prime-k$. 
We then employ a voting mechanism on these potential origin indices to determine the final matching region. 

\subsection{Change Detector} \label{sec:Change Detector}




\subsubsection{Ensemble Differencing} \label{sec:Ensemble Differencing}
\noindent
For the candidate image $\mathrm{I}$ and the selected reference $\mathrm{I_{ref}}$, we compute the difference between these two image matrices using $\Delta \mathrm{I}=\mathrm{I}-\mathrm{I_{ref}}$.
The resulting differencing matrix $\Delta \mathrm{I}$ highlights the pixel discrepancies between these two images, i.e., the RoIs in the satellite image during the revisiting interval. 
Ideally, the output should be a sparse differencing matrix, where most elements are zeros. 
However, even when comparing two highly similar images, the subtraction can yield a differencing matrix with numerous pixel value locations (i.e., dense matrix), potentially leading to a large output size. 
This discrepancy in size can be attributed to prevalent image noise, arising from optical hardware imperfections and environmental variations. 
Such subtle pixel perturbations, i.e., RoNs, can significantly reduce the effectiveness of the ensemble differencing process.

\begin{figure}[t]
    \centering
    \includegraphics[width=0.48\textwidth]{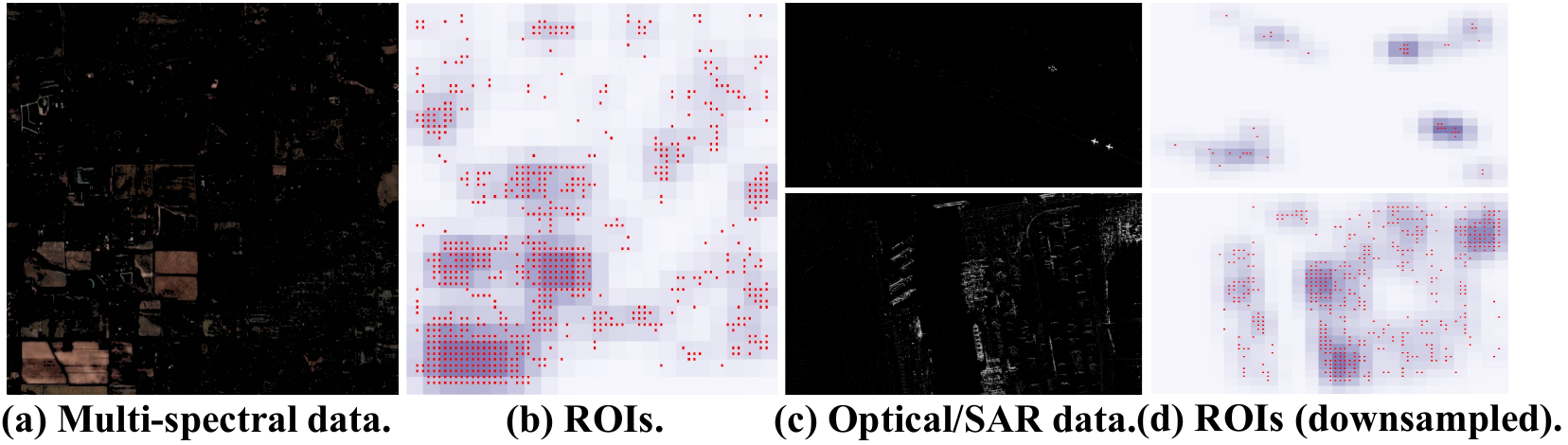}
    \vspace{-0.05in}
    \caption{Illustration of the change detector on various data modalities, including multi-spectral, optical, and Synthetic Aperture Radar (SAR).}
   \vspace{-0.1in}
    \label{fig:Differencing}
\end{figure}

\subsubsection{Local Compensation} \label{sec:Local Compensation} 
\textbf{RoN identification.} 
Regarding such a dense differencing matrix, we propose a local compensation method. 
Exploiting the similarity between layers in an image, we capitalize on the ``one-to-many'' notion that pixel-point variations across layers are akin. 
This allows us to pinpoint noise points in one single channel to identify RoNs, thereby enhancing computational efficiency. 
Specifically, focusing on the first channel (typically the Red channel) of the differencing matrix $\Delta \mathrm{I}_{(w,h,0)}$, we traverse all pixel values for RoNs detection. 
Noise introduces subtle perturbations in pixel values, which tend to be near zero but not exactly zero. 
These perturbations, i.e., RoNs, can be identified by comparing the pixel values against a predefined local compensation threshold $\tau$. 
We then retain the original values for RoIs, while setting the values in RoNs to zero via 
\begin{equation}
\footnotesize
\Delta \mathcal{I}_{(h,w,n)}= \begin{cases} \Delta \mathcal{I}_{(h,w,n)}, & \text{if } \Delta 
 \mathcal{I}_{(h,w,0)}> max(\mathcal{I}_{(h,w,0)}) \times \tau \\ 
0, & \text{otherwise. } \end{cases} 
\end{equation}

\noindent
\textbf{Sparse representation.} 
Following ensemble differencing and local compensation, a sparse matrix of RoIs is obtained, with most values being zero, as illustrated in~\autoref{fig:Differencing}.
To optimize its size for transmission, we encode it in Compressed Sparse Row (CSR)~\cite{greathouse2014efficient} format. 
CSR comprises three components: non-zero element value array $V$, the column index array $\text{Col\_index}$, and the row pointer array $\text{Row\_ptr}$. 
For a sparse matrix $\Delta \mathrm{I}^{(H, W, N)}$ with a total of $NZ$ non-zero elements, we first reshape it into $\Delta \mathrm{I}^{\prime(H \times W, \: N)}$ then apply CSR encoding. 
With these three arrays, the CSR representation of the sparse matrix $\Delta \mathrm{I}^{\prime(H \times W, \: N)}$ can be represented by
\begin{equation}
\small
\Delta \mathrm{I}^{\prime}_{(i,j)}=
\begin{cases} 
V[k], & \text{if } c=\text{col\_index}[k] \: \&\&   \: r_i \leq k < r_{i+1} \\
0, & \text{otherwise,}
\end{cases}
\end{equation}
where $i$ and $j$ represent the indices of the row and column of $\Delta \mathrm{I}^{\prime}$, respectively. 


%


\begin{figure}[t]
    \centering
    \includegraphics[width=0.47\textwidth]{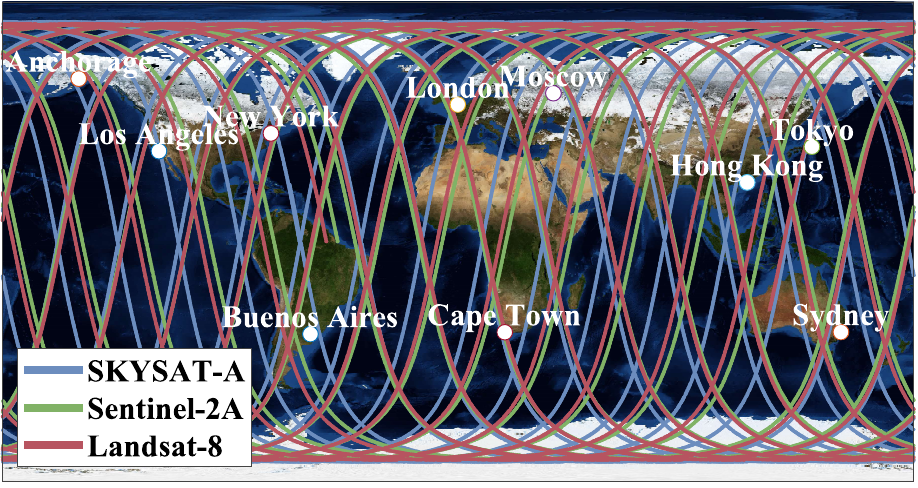}
    \vspace{-0.1in}
    \caption{Illustration of satellite tracks and ground station locations on the Earth map under Mercator projection.}
   \vspace{-0.1in}
    \label{fig:Satellite_System}
\end{figure}

\section{Methodology} \label{sec:Methodology}

\noindent
\textbf{Software.}
We implement \System as a Python-based processing pipeline that operates in two stages: (1) satellites capture raw observations and transform them into intermediate representations, and (2) ground stations receive and reconstruct the original data.

\noindent
\textbf{Satellite and ground compute systems.}
We deploy our applications on NVIDIA Jetson TX2 as the OBC platform, as the commodity computing devices (e.g., NVIDIA Jetson TX2/AGX Orin mobile GPUs) are compatible with being deployed in LEO satellites~\cite{denby2020orbital}. 
Specifically, Jetson TX2 is built with an 256-core NVIDIA Pascal$^\text{TM}$-family GPU architecture, featuring \SI{8}{GB} memory, and operating in 7.5/\SI{15}{W} power modes. 
The ground station is equipped with a workstation (with Intel Core i7-10700 \SI{2.90}{GHz} CPU, \SI{64}{GB} RAM, and NVIDIA GeForce RTX 3080 GPU).

\noindent
\textbf{Satellite constellation simulator.}
We evaluate \System using the simulator that models orbital dynamics, data acquisition, and satellite communication behaviors.
The simulator leverages real-world TLE orbit descriptors from CelesTrak to describe the orbital characteristics of Earth-orbiting satellites.
As illustrated in~\autoref{fig:Satellite_System}, we consider three representative Earth observation satellites: Landsat-8, Sentinel-2A, and SKYSAT-A, operating during one revisiting cycle (i.e., 24 hours). 
Additionally, ten ground stations are evenly distributed across the globe.

\begin{figure}[t]
\vspace{-0.15in}
\centering
     \subfloat[Flat-Sat platform.]{\includegraphics[width=0.28\textwidth]{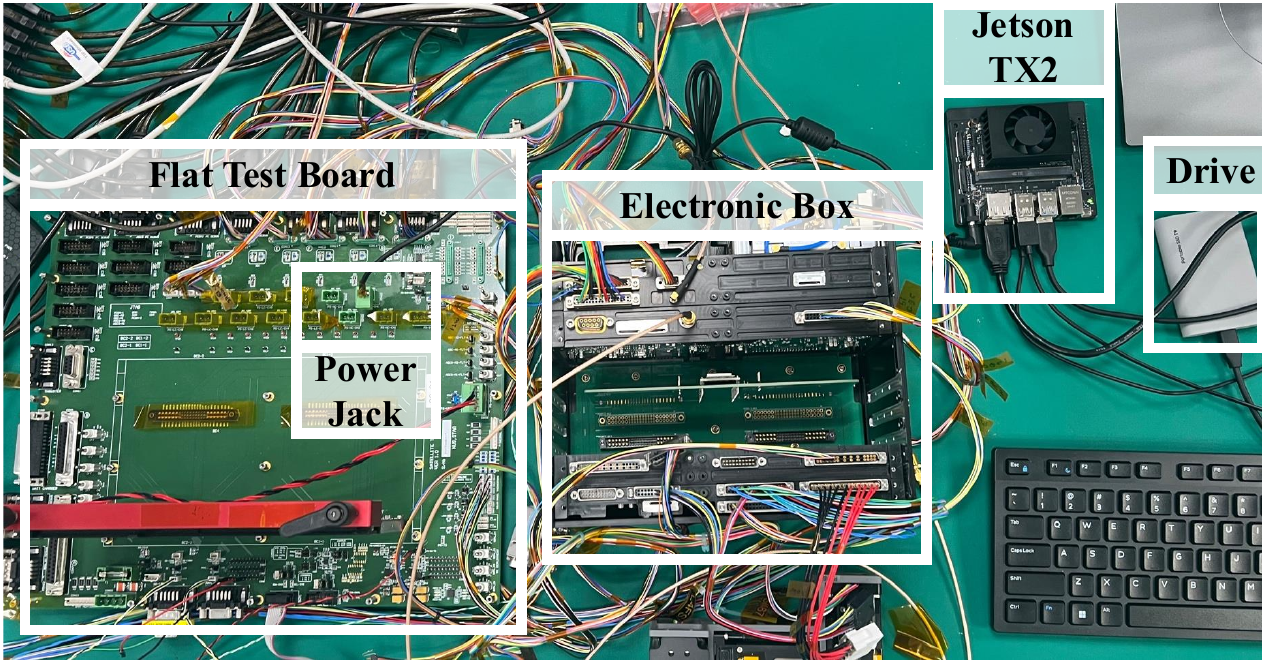} \label{fig:Flat_Sat}} 
     \subfloat[Anonymous flight model.]{\includegraphics[width=0.205\textwidth]{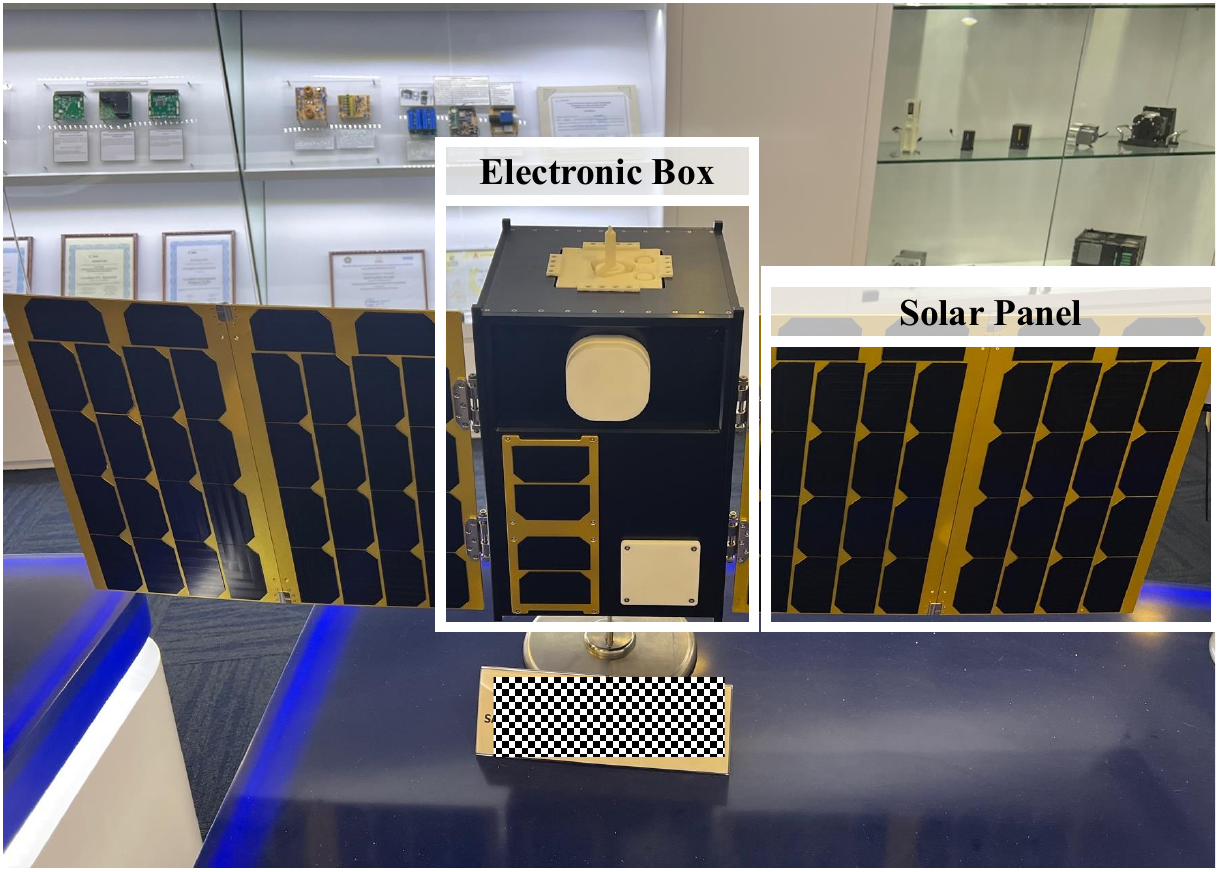} \label{fig:LUMELITE}} 
     \vspace{-0.1in}
\caption{Flat-Sat deployment.}
\vspace{-0.05in}
\label{fig:Flat_Sat_Setup}
\end{figure}

\noindent
\textbf{Flat-Sat testbed.}
To quantify the system overhead, we evaluate \System on a Flat-Sat testbed (see~\cref{fig:Flat_Sat}), a bench-top satellite prototype integrating key flight subsystems, including the OBC, communication interface module (CIM), attitude determination and control system (ADCS), and electrical power system (EPS).
These components are interconnected using flight-equivalent harnessing.
The Flat-Sat is derived from an already launched satellite platform; a representative replica is shown in~\cref{fig:LUMELITE}. 
For anonymity, we omit identifying details of the original flight model.

\noindent
\textbf{Applications.}
As illustrated in~\autoref{tab:dataset}, we evaluate \System across five popular applications. 



\begin{table}[!t]
\vspace{-0.1in}
\centering
\caption{Statistics of application datasets.}
\vspace{-0.05in}
\label{tab:dataset}
\renewcommand{\arraystretch}{1.2}
\resizebox{0.46\textwidth}{!}{%
\begin{tabular}{lccccccl}
\toprule[1.5pt]
\textbf{Dataset} &
  \textbf{\begin{tabular}[c]{@{}l@{}}\#Image/\\      Frame\end{tabular}} &
  \textbf{\#AOIs} &
  \textbf{\begin{tabular}[c]{@{}l@{}}Mean \\      GSD (m)\end{tabular}} &
  \textbf{\begin{tabular}[c]{@{}l@{}}Temporal \\      Resolution\end{tabular}} &
  \textbf{\begin{tabular}[c]{@{}l@{}}Time\\      Span\end{tabular}} &
  \textbf{Source} \\ \hline
\rowcolor[HTML]{EFEFEF} 
\textit{SpaceNet 7}~\cite{van2021spacenet}               & 1,423   & 101 & 4   & Monthly      & 2 years & PlanetScope                         \\
\textit{LSCIDMR}~\cite{bai2021lscidmr}                   & 104,390 & 11  & -   & Daily        & 1 year & Himawari-8                           \\
\rowcolor[HTML]{EFEFEF} 
\textit{DynamicEarthNet}~\cite{toker2022dynamicearthnet} & 54,750  & 75  & 3   & Daily        & 2 years & PlanetFusion                        \\
\textit{MTGL40-5}~\cite{ma2023mtgl40}                    & -       & 40  & 0.5 & Yearly       & 5 years & {\color[HTML]{222222} Google Earth} \\
\rowcolor[HTML]{EFEFEF} 
\textit{SatSOT}~\cite{zhao2022satsot}                    & 27,664  & 105 & -   & Sub-secondly & -      & Skybox,  etc.       \\     
\bottomrule[1.5pt]
\end{tabular}%
}
\begin{tablenotes}
\item \textit{(AOI: area of interest, GSD: ground sampling distance).}
\end{tablenotes}
\vspace{-0.1in}
\end{table}

\begin{figure*}[t]
\centering
\vspace{-0.2in}
    \subfloat[Transmission load.]{\includegraphics[width=0.235\textwidth]{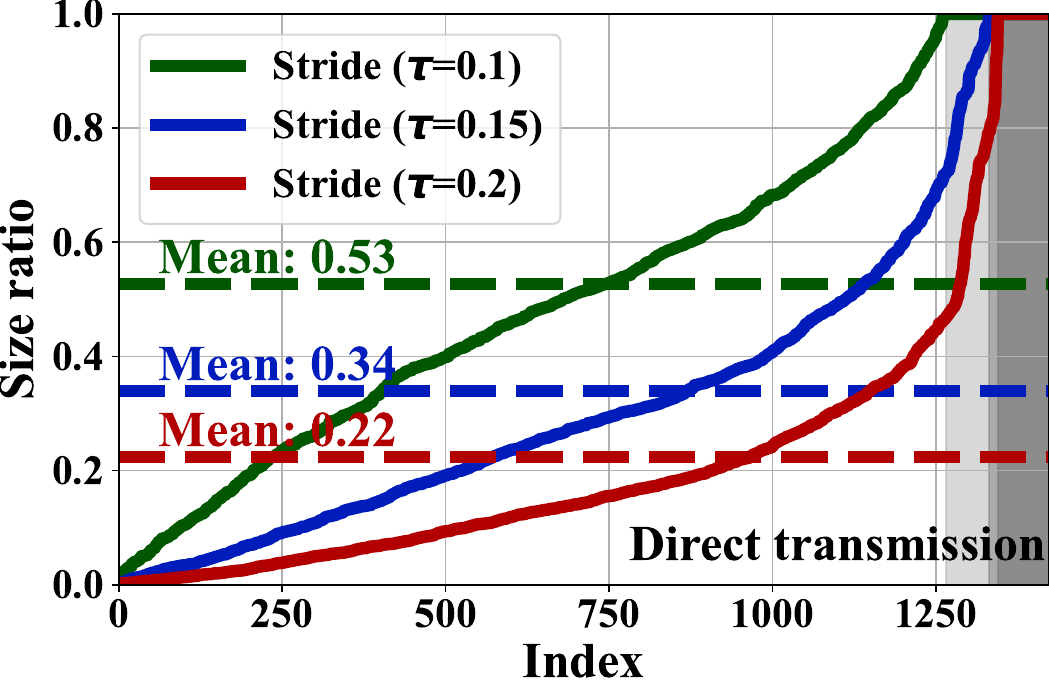}  \label{fig:Overall_Performance1}}
    \subfloat[Size ratio distribution.]{\includegraphics[width=0.245\textwidth]{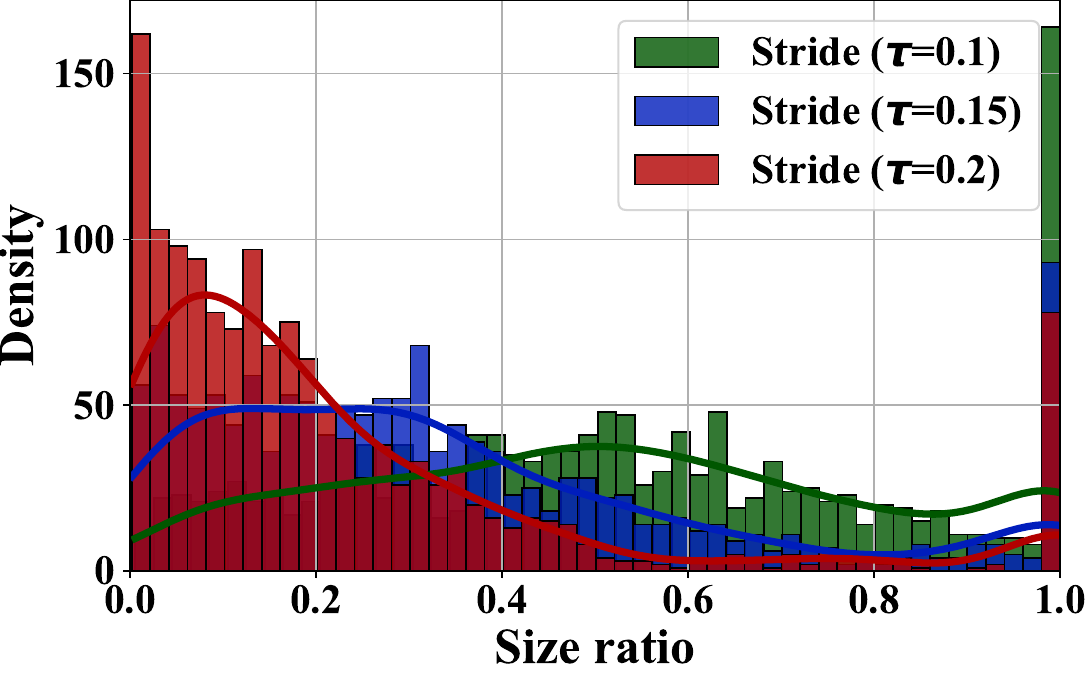}  \label{fig:Overall_Performance1_1}}
    \subfloat[Reconstruction quality.]{\includegraphics[width=0.255\textwidth]{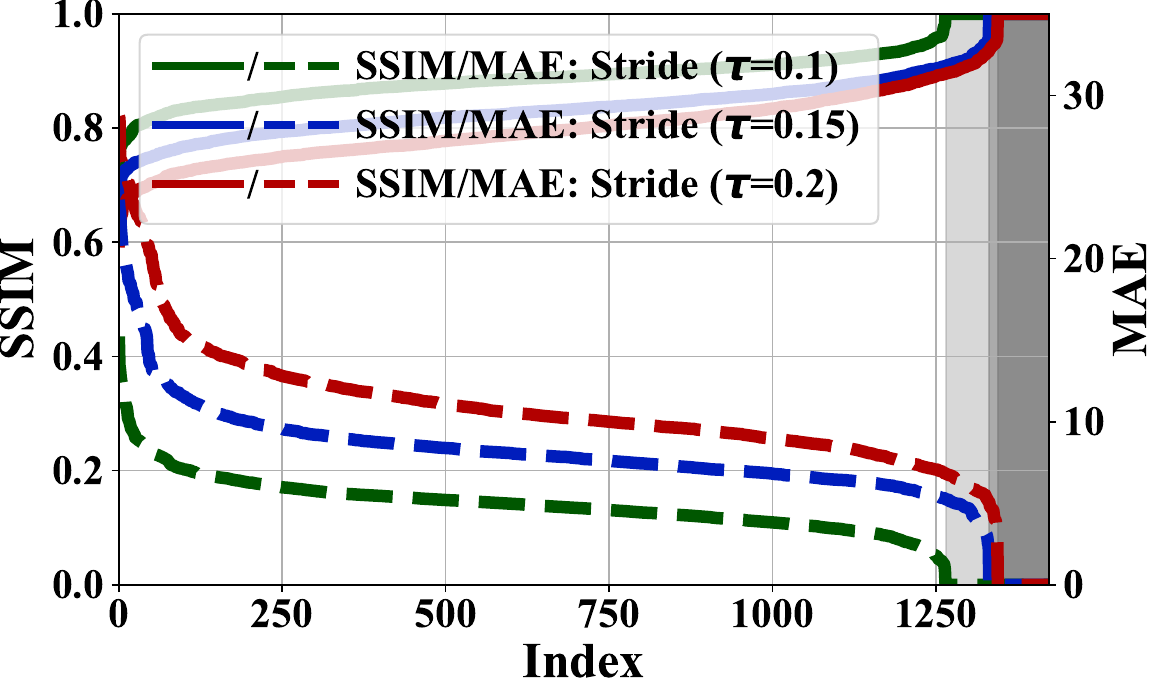}  \label{fig:Overall_Performance2}}
    \subfloat[SSIM distribution.]{\includegraphics[width=0.245\textwidth]{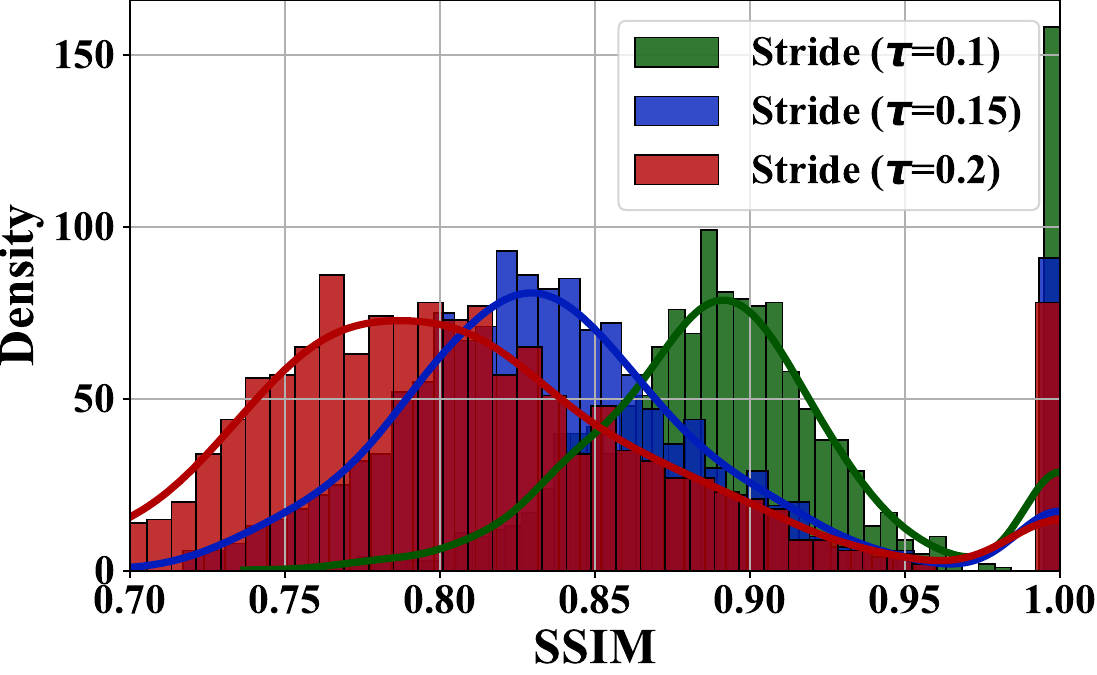}  \label{fig:Overall_Performance2_1}}
    \vspace{-0.05in}
    \caption{Overall performance of \System in terms of transmission load and reconstruction quality.}
    \label{fig:Overall_Performance}
\vspace{-0.05in}
\end{figure*}


\begin{figure*}[t]
\vspace{-0.15in}
\centering
\begin{minipage}{0.485\linewidth}
\centering
\subfloat[Revisiting imagery delivery.]{\includegraphics[width=0.5\textwidth]{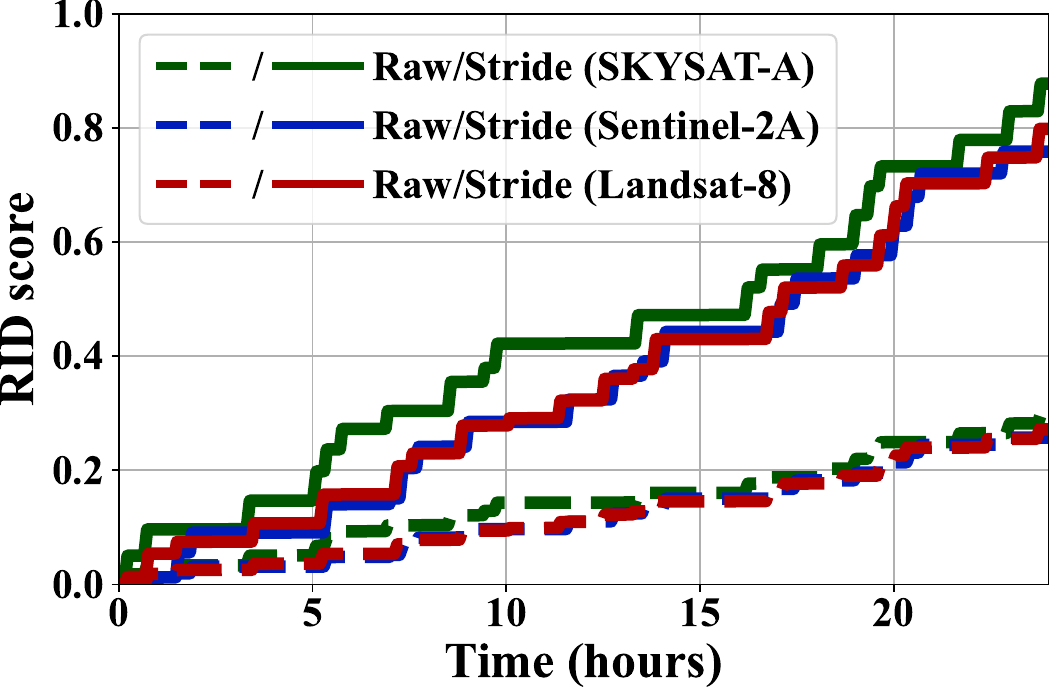}  \label{fig:Overall_Performance3}}
\subfloat[CDF of RID scores.]{\includegraphics[width=0.5\textwidth]{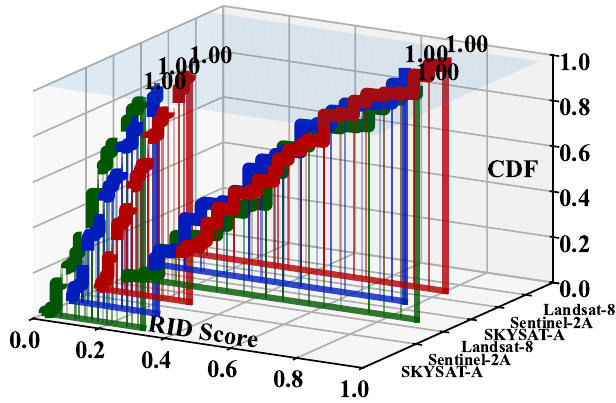}  \label{fig:Overall_Performance3_1}}
\vspace{-0.05in}
\caption{Overall performance in revisiting capability.}
\vspace{-0.1in}
\label{fig:Overall_Performance_}
\end{minipage}
\begin{minipage}{0.49\linewidth}
\centering
\subfloat[Accuracy.]{\includegraphics[width=0.5\textwidth]{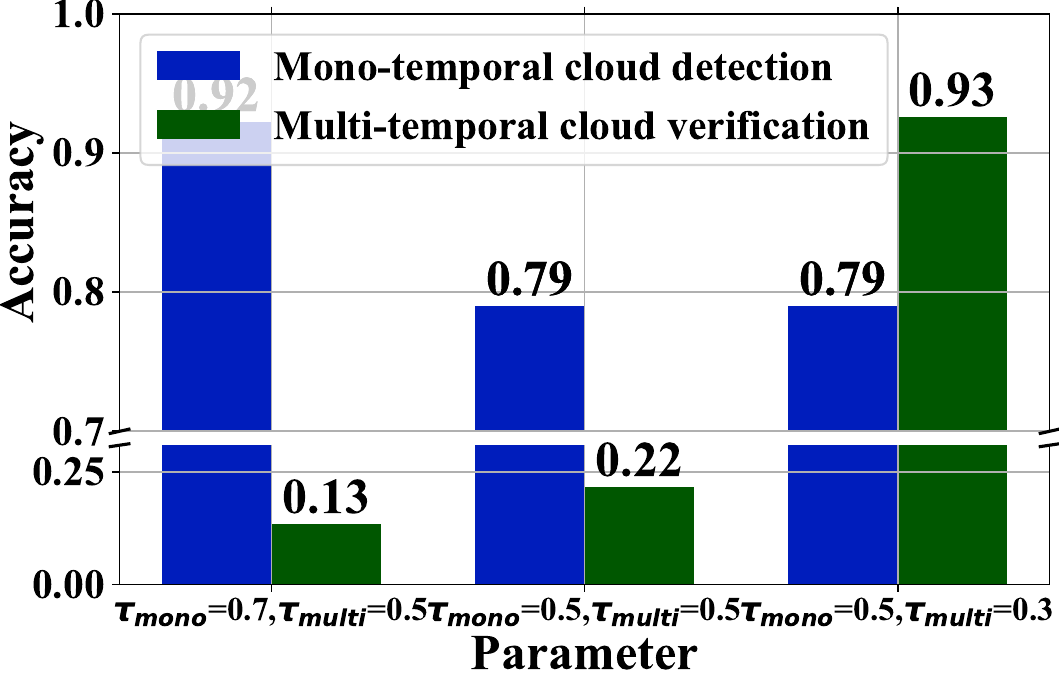}  \label{fig:Cloud_Indicator1}}
\subfloat[Accuracy variation.]{\includegraphics[width=0.5\textwidth]{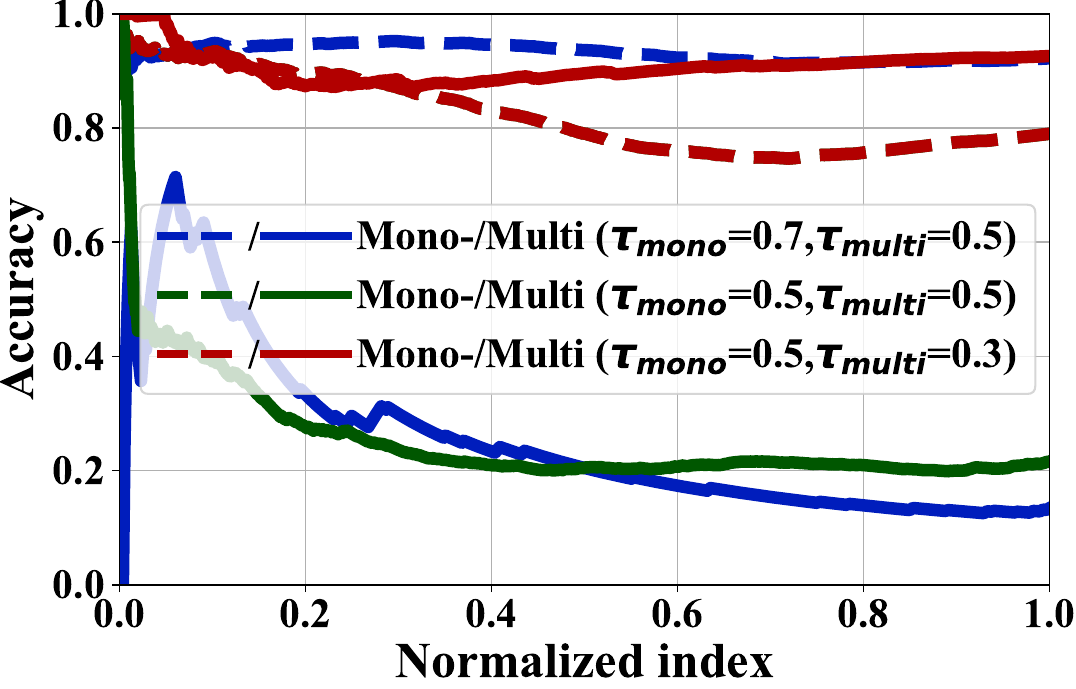}  \label{fig:Cloud_Indicator2}}
\vspace{-0.05in}
\caption{Results of the cloud indicator.}
\vspace{-0.1in}
\label{fig:Cloud_Indicator}
\end{minipage}
\end{figure*}

\noindent
\textbf{Default parameters.} 
For the cloud indicator, \System utilizes 3 bands, with thresholds $\tau_{mono} = 0.7$ and $\tau_{multi} = 0.7$. 
For the reference selector, the algorithm parameters are $\tau_\lambda = \tau_\phi = 0.01$ and $\epsilon = 1\times10^7$. 
For the change detector, the local compensation parameter $\tau$ is set to 0.15.

\noindent
\textbf{Benchmarks.}
We compare \System with four baselines:
\textbf{(1) \Raw}, a plain Earth observation satellite system without data compression; 
\textbf{(2) JPEG-2000~\cite{taubman2002jpeg2000}}, a conventional imagery compression method; 
\textbf{(3) CCSDS 121.0-B-3~\cite{CCSDS121B3}}, an imagery compression standard made by the Consultative Committee for Space Data Systems (CCSDS). 
\textbf{(4) Earth+~\cite{du2025earth+}}, the recent reference uplinking-based compression method. 


\noindent
\textbf{Metrics.}
We consider the following performance aspects: 
\textbf{(1) Transmission load.} We utilize the size ratio (of the original images) to measure the transmission load of the satellite downlink transmission. 
\textbf{(2) Reconstruction quality.} We employ the Structural Similarity Index Measure (SSIM) and Mean Absolute Error (MAE) to assess the quality of the reconstructed images. 
\textbf{(3) Revisiting capability.} We utilize the RID score to describe the revisiting capability of satellites. For fair comparison across different satellites, RID score is defined as the ratio of successfully delivered data to the total collected data within a fixed time window (i.e., 24 hours).


\section{Evaluation} \label{sec:Evaluation}
\subsection{Overall Performance} \label{sec:Overall Performance}
We evaluate the overall performance of \System on the \textit{SpaceNet 7}~\cite{van2021spacenet} dataset. 
\cref{fig:Overall_Performance1} illustrates the transmission load of \System with three different local compensation parameters $\tau$. 
Overall, \System demonstrates a significant reduction in size ratio, which means the transmitted data volume is much lower than the original imagery. 
Specifically, \System attains size ratios of 0.22, 0.34, and 0.53 for $\tau$ of 0.2, 0.15, and 0.1, respectively. 
The observed trend can be attributed to the fact that increasing the local compensation parameter $\tau$ tends to enlarge RoNs, thereby consistently decreasing the size ratios. 
This is evident in~\cref{fig:Overall_Performance1_1}, the density distribution of the size ratio. 
Additionally, the shadow region indicates direct transmission, including cloud images, reference images, and overloaded data. 



\cref{fig:Overall_Performance2} illustrates the reconstruction quality of \System on ground stations. 
Overall, \System consistently maintains a relatively high SSIM and low pixel-level MAE, which verifies the effectiveness of \System. 
Likewise, with the increment of local compensation parameter $\tau$, there is a slight decrease in reconstruction quality due to the enlarged RoNs (as evident in~\cref{fig:Overall_Performance2_1}). 


\cref{fig:Overall_Performance3} and~\cref{fig:Overall_Performance3_1} illustrate the revisiting imagery delivery performance of \System for three different Earth observation satellites under $\tau$=0.15. 
We observe that \System significantly improves the RID score across various types of Earth observation satellites. 
For the SKYSAT-A satellite, \System attains an RID score of 0.8776, compared to 0.2982 for \Raw, resulting in an improvement of revisiting imagery delivery by 2.94$\times$. 
Likewise, \System achieves an improvement of RID score by 1.89$\times$, and 4.55$\times$ for $\tau=$ 0.1 and 0.2, respectively. 
This is primarily attributed to the reduction in transmission load, which allows for more efficient revisiting imagery delivery.

\begin{figure*}
\centering
\begin{minipage}{0.49\linewidth}
    \centering
	\subfloat[Voting origin accuracy.]{\includegraphics[width=0.525\textwidth]{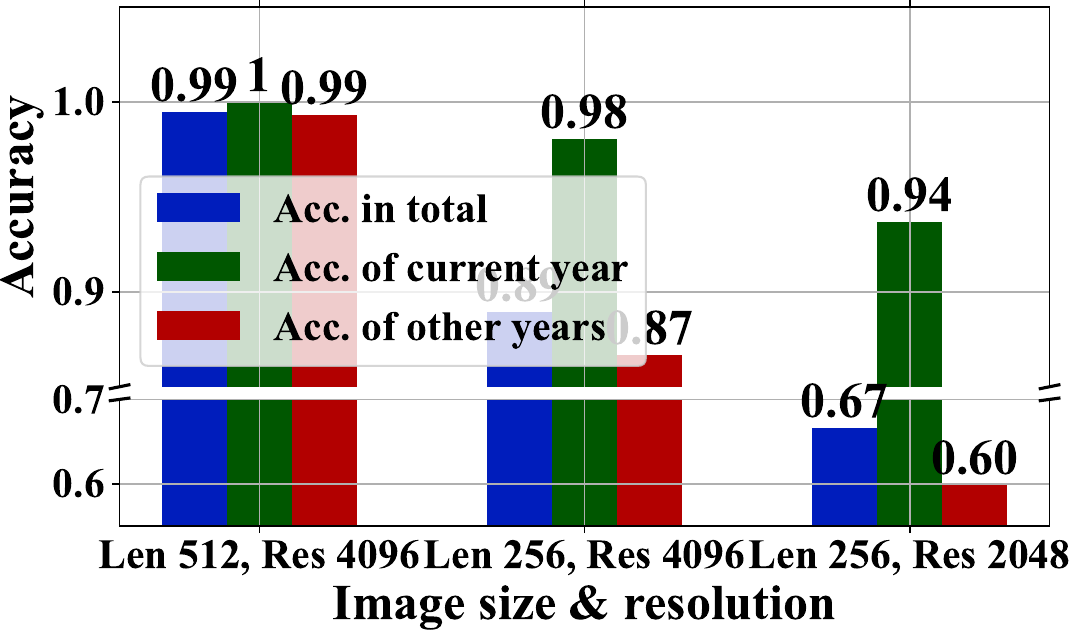} \label{fig:Reference_Selector1}}
	\subfloat[Matching region MAE.]{\includegraphics[width=0.475\textwidth]{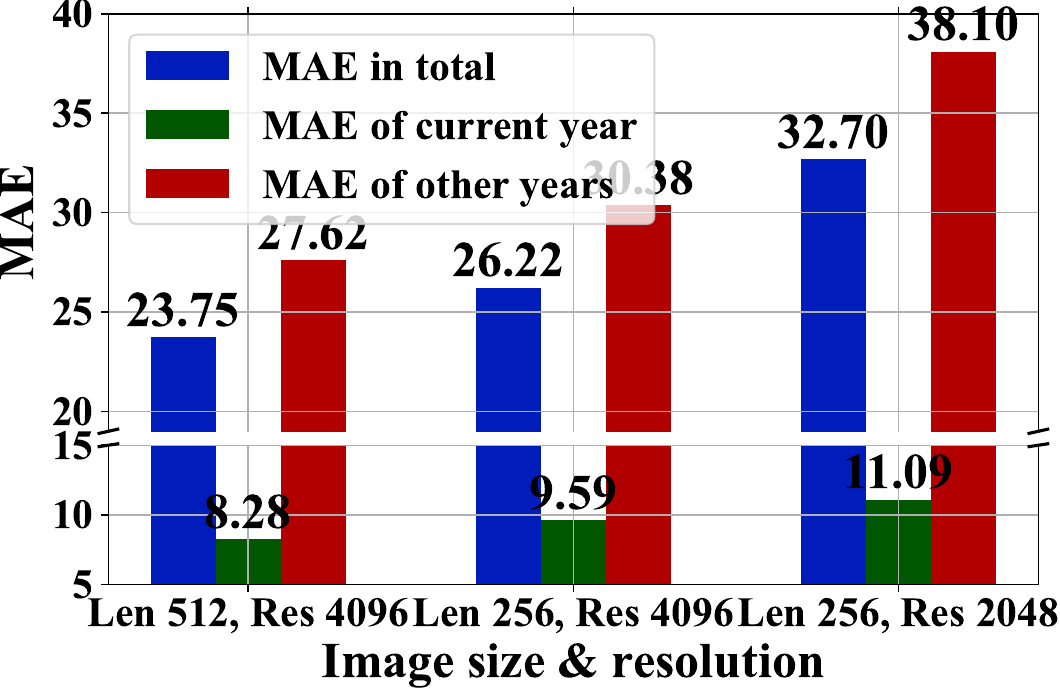} \label{fig:Reference_Selector2}}
\vspace{-0.05in}
\caption{Results of the reference selector.}
\vspace{-0.1in}
\label{fig:Reference_Selector}
\end{minipage}
\begin{minipage}{0.49\linewidth}
    \centering
	\subfloat[Transmission load over time.]{\includegraphics[width=0.48\textwidth]{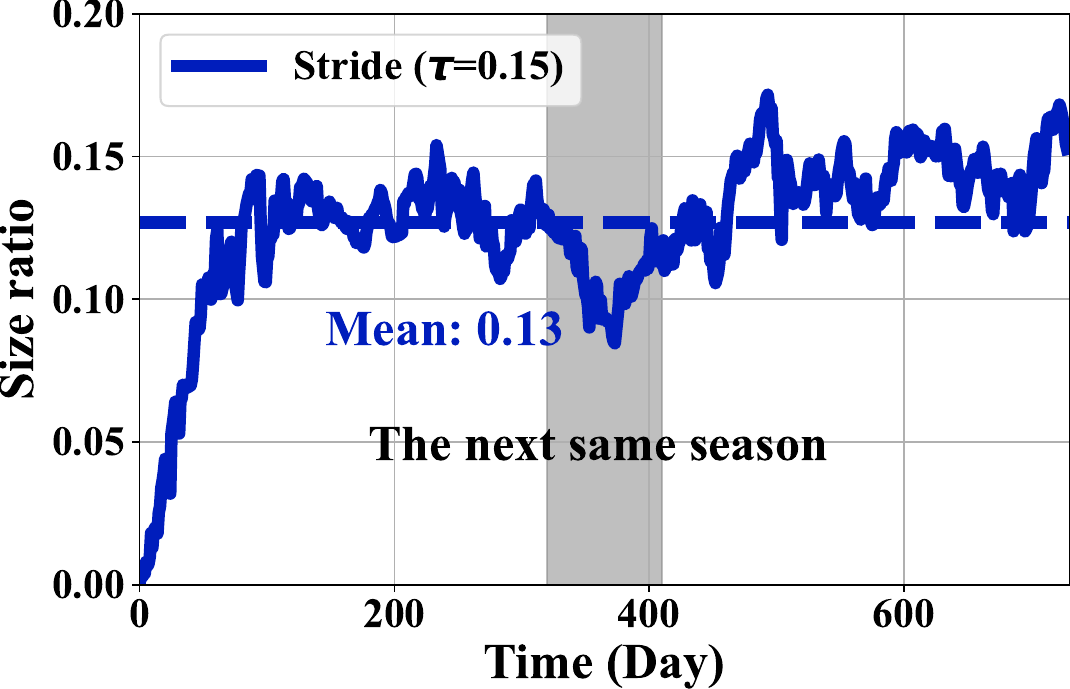} \label{fig:Change_Detector1}}
	\subfloat[Reconstruction quality.]{\includegraphics[width=0.52\textwidth]{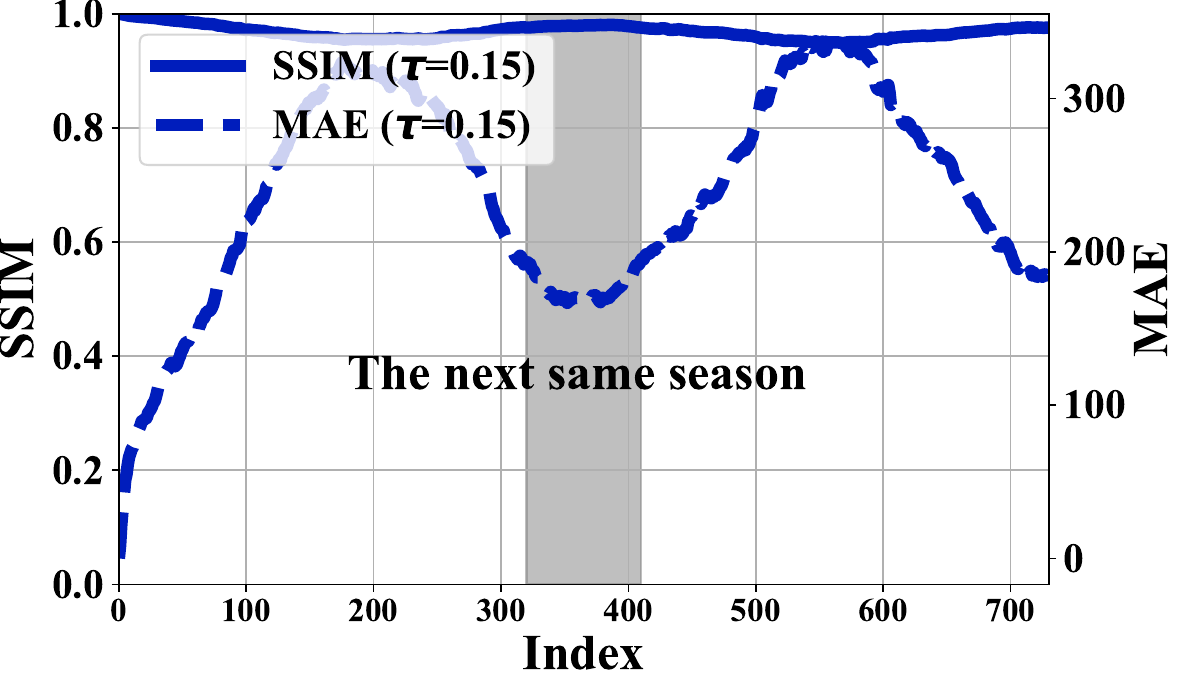} \label{fig:Change_Detector2}}
\vspace{-0.05in}
\caption{Results of the change detector.}
\vspace{-0.1in}
\label{fig:Change_Detector}
\end{minipage}
\end{figure*}

\begin{figure*}[t]
\vspace{-0.15in}
\centering
     \subfloat[RAM usage.]{\includegraphics[width=0.24\textwidth]{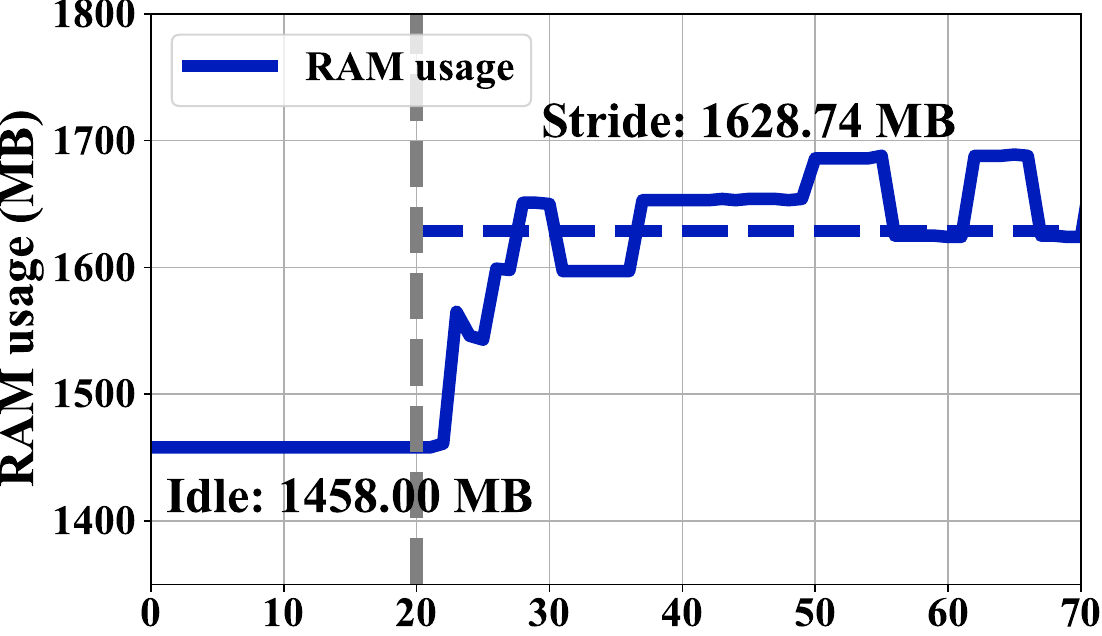} \label{fig:System_Overhead1}}
     \subfloat[CPU usage.]{\includegraphics[width=0.24\textwidth]{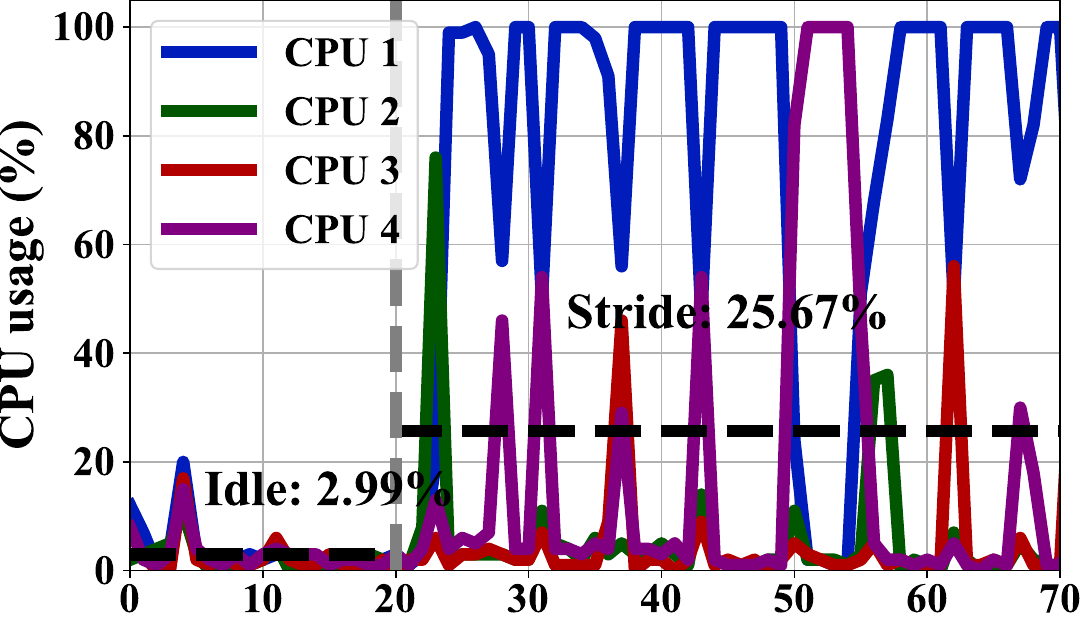} \label{fig:System_Overhead2}}
     \subfloat[Temperature.]{\includegraphics[width=0.24\textwidth]{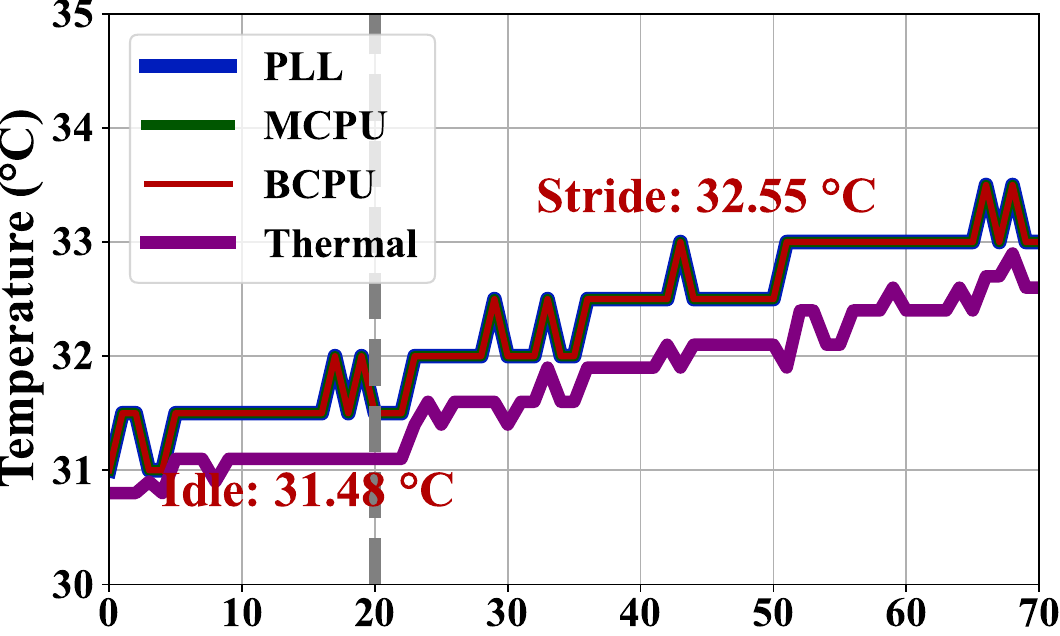} \label{fig:System_Overhead3}}
     \subfloat[Energy consumption.]{\includegraphics[width=0.24\textwidth]{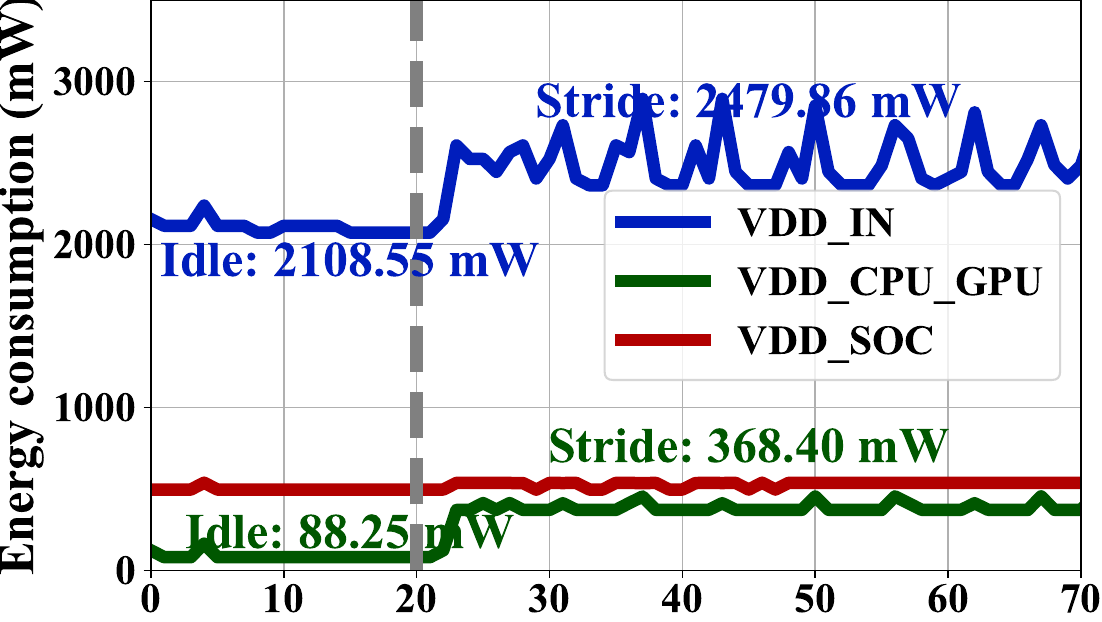} \label{fig:System_Overhead4}}
\vspace{-0.05in}
\caption{System overhead of \System in terms of RAM usage, CPU load, temperature, and energy consumption.}
\vspace{-0.1in}
\label{fig:System_Overhead}
\end{figure*}

\subsection{Evaluation of Cloud Indicator} \label{sec:Evaluation of Cloud Indicator}
We evaluate \System's cloud indicator on the \textit{LSCIDMR}~\cite{bai2021lscidmr} dataset under different $\tau_{mono}$ and $\tau_{multi}$.
As illustrated in~\cref{fig:Cloud_Indicator}, a larger $\tau_{mono}$ yields superior performance for a fixed $\tau_{multi}$, since a smaller $\tau_{mono}$ tends to misclassify objects with relatively high pixel values.
Conversely, for a fixed $\tau_{mono}$, a smaller $\tau_{multi}$ performs better, likely due to increased tolerance for variations in spatial extents across multi-temporal images.
Moreover, $\tau_{mono}$ also affects multi-temporal cloud verification by shaping the cloud cache.
Overall, the cloud indicator is satisfactory: accurate mono-temporal detection reduces computational load, while multi-temporal verification still effectively prevents the direct transmission of part of non-cloudy images.

\subsection{Evaluation of Reference Selector} \label{sec:Evaluation of Reference Selector}
We evaluate \System's reference selector on the \textit{MTGL40-5} dataset across capture years and image attributes.
We concatenate images to create a larger query image set and exclude images containing large black regions.
A voting origin is accurate if its distance from the ground truth is below a small fraction of the maximum pixel value.
\cref{fig:Reference_Selector1} shows that \System achieves high voting origin accuracy overall.
\System benefits most from the most recent capture year, as older images tend to have undergone more changes that reduce alignment effectiveness, and performance further improves with larger image sizes and higher resolutions.
\cref{fig:Reference_Selector2} confirms these trends via MAE, which is inversely related to accuracy.
Fortunately, satellite imagery typically satisfies the large-size and high-resolution criteria, so these limitations are generally not a concern.

\subsection{Evaluation of Change Detector} \label{sec:Evaluation of Change Detector}
We evaluate \System's change detector on the \textit{DynamicEarthNet}~\cite{toker2022dynamicearthnet} dataset over a two-year period.
\cref{fig:Change_Detector1} shows that \System substantially reduces the transmission size ratio.
The size ratio indirectly reflects reference similarity, exhibiting seasonal minima in the shaded regions, as imagery captures similar landscapes within the same season.
\cref{fig:Change_Detector2} further shows that \System consistently maintains high reconstruction quality in both SSIM and MAE.

\subsection{System Overhead on Flat-Sat Platform} \label{sec:System Overhead on Flat-Sat Platform}
\autoref{fig:System_Overhead} illustrates the system overhead of Jetson TX2 (i.e., OBC) running \System on the Flat-Sat platform, focusing on four key metrics, i.e., RAM usage, CPU load, temperature, and energy consumption.
\cref{fig:System_Overhead1} shows that \System introduces only a modest RAM increase over the idle state, leaving a large portion of the available memory free and demonstrating efficient memory utilization under the given workload.
\cref{fig:System_Overhead2} presents the CPU load across the four cores relative to their operating frequency, where the mean CPU usage of \System remains at a moderate level, reflecting the processor's efficiency under workload conditions.
\cref{fig:System_Overhead3} shows the temperature variations of four key components: the Phase-Locked Loop (PLL), the Microcontroller CPU (MCPU), the Board CPU (BCPU), and the entire hardware system (Thermal).
As expected, their temperatures gradually increase with operation time and eventually stabilize within a reasonable and acceptable thermal range, indicating efficient heat dissipation and safe operating conditions for \System.
\cref{fig:System_Overhead4} illustrates the energy consumption of three components.
The total power consumption (VDD\_IN) and the CPU and GPU energy consumption (VDD\_CPU\_GPU) exhibit only slight and consistent increases, while the System-On-Chip energy consumption (VDD\_SOC) stabilizes, suggesting that its power usage remains steady despite fluctuations in other components.
These metrics demonstrate the low system overhead of \System, underscoring its suitability for power-constrained devices.

\begin{figure*}[!t]
\vspace{-0.15in}
\centering
	\subfloat[Transmission load. ]{\includegraphics[width=0.245\textwidth]{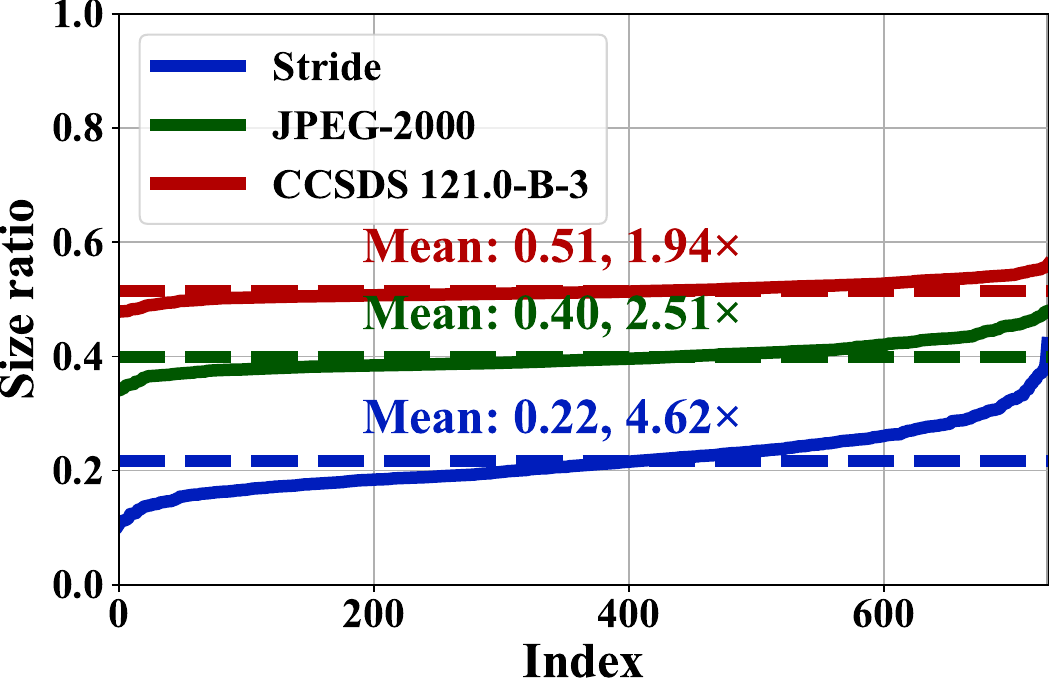}  \label{fig:Comparison1}}
	\subfloat[Density distribution.]{\includegraphics[width=0.245\textwidth]{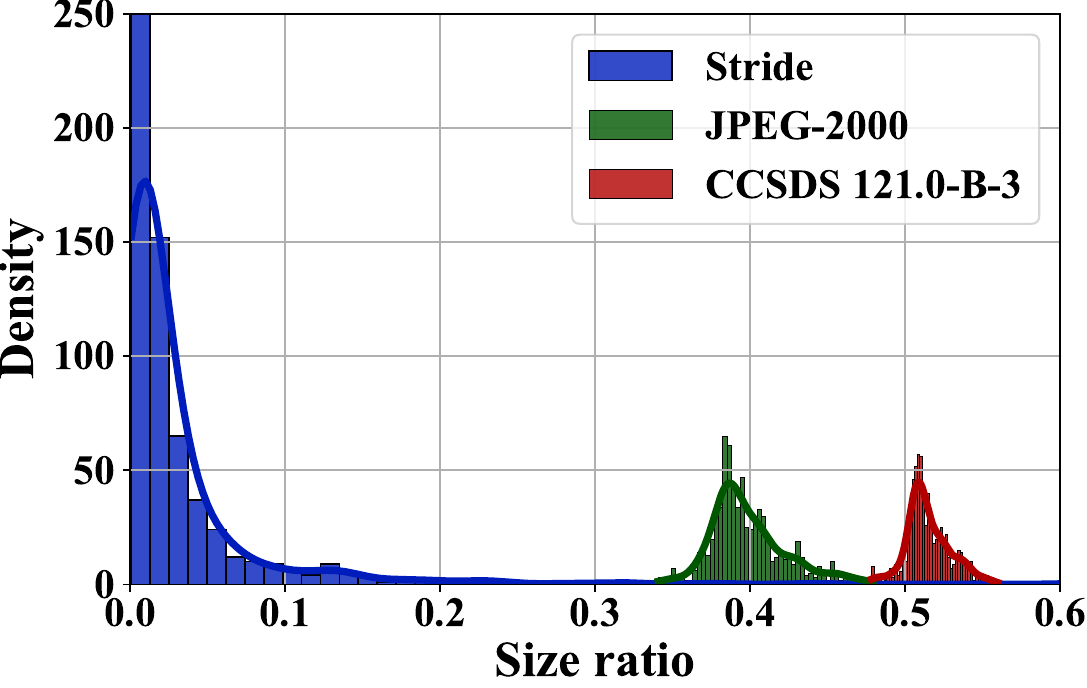}  \label{fig:Comparison2}}
    \subfloat[Transmission load. ]{\includegraphics[width=0.245\textwidth]{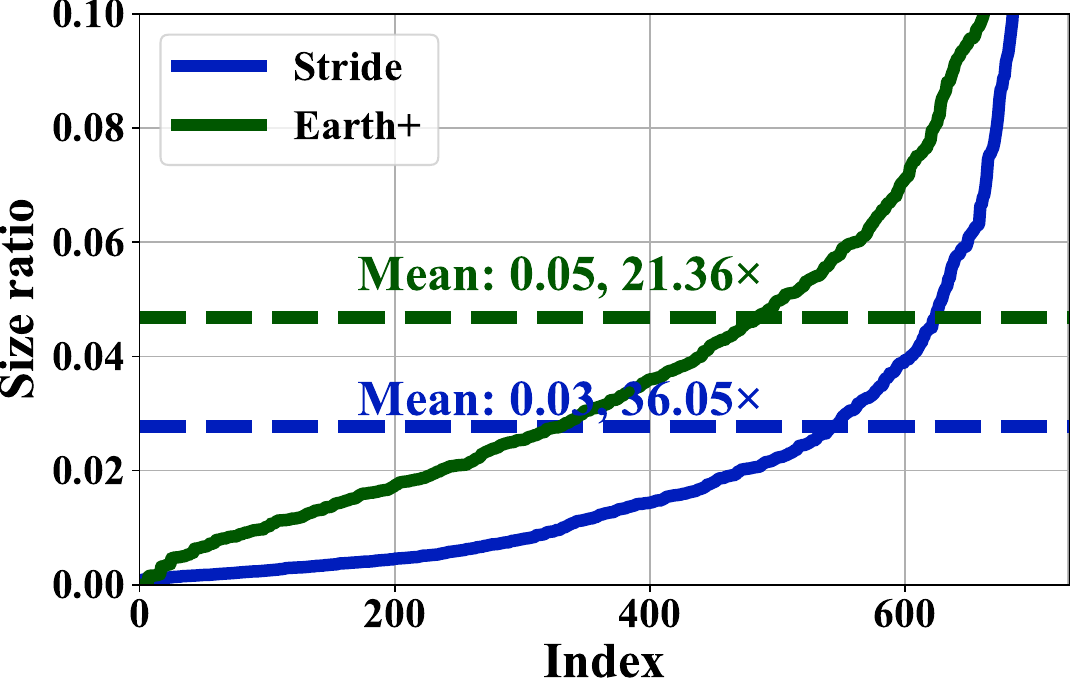}  \label{fig:Comparison3}}
	\subfloat[Density distribution.]{\includegraphics[width=0.245\textwidth]{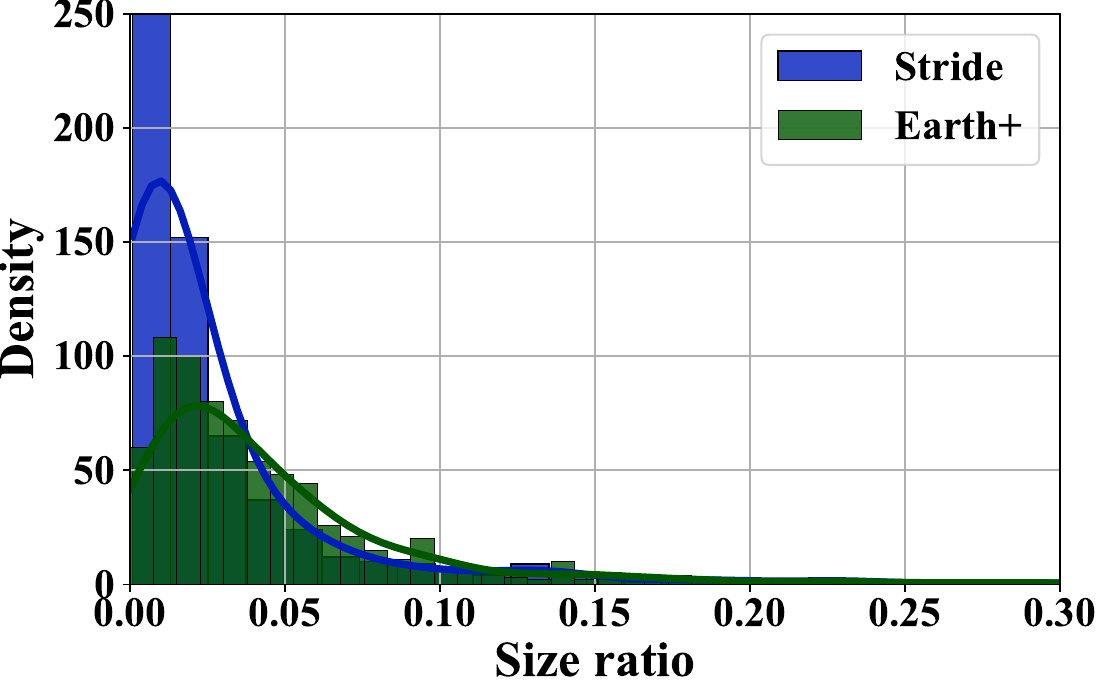}  \label{fig:Comparison4}}
\vspace{-0.05in}
\caption{Quantitative comparison with benchmarks (lossless: (a)-(b), lossy: (c)-(d)).}
\vspace{-0.2in}
\label{fig:Comparison}
\end{figure*}

\begin{figure*}
\begin{minipage}{0.5\linewidth}
\centering
	\subfloat[Transmission load over frames.]{\includegraphics[width=0.5\textwidth]{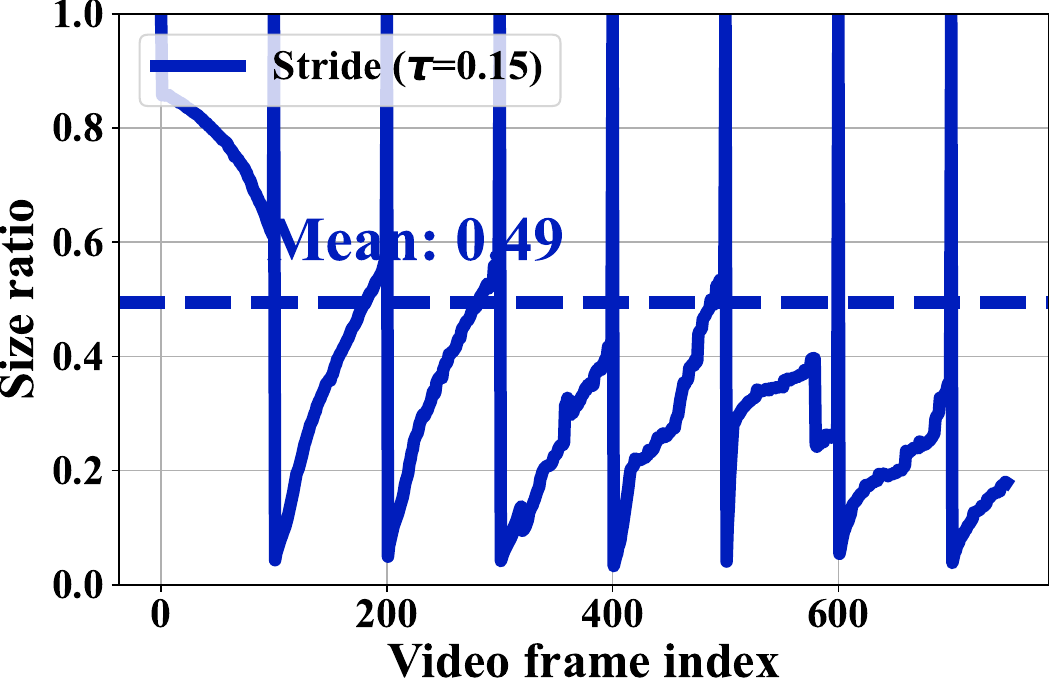}  \label{fig:Use_Case1_1}}
	\subfloat[Revisiting imagery delivery.]{\includegraphics[width=0.5\textwidth]{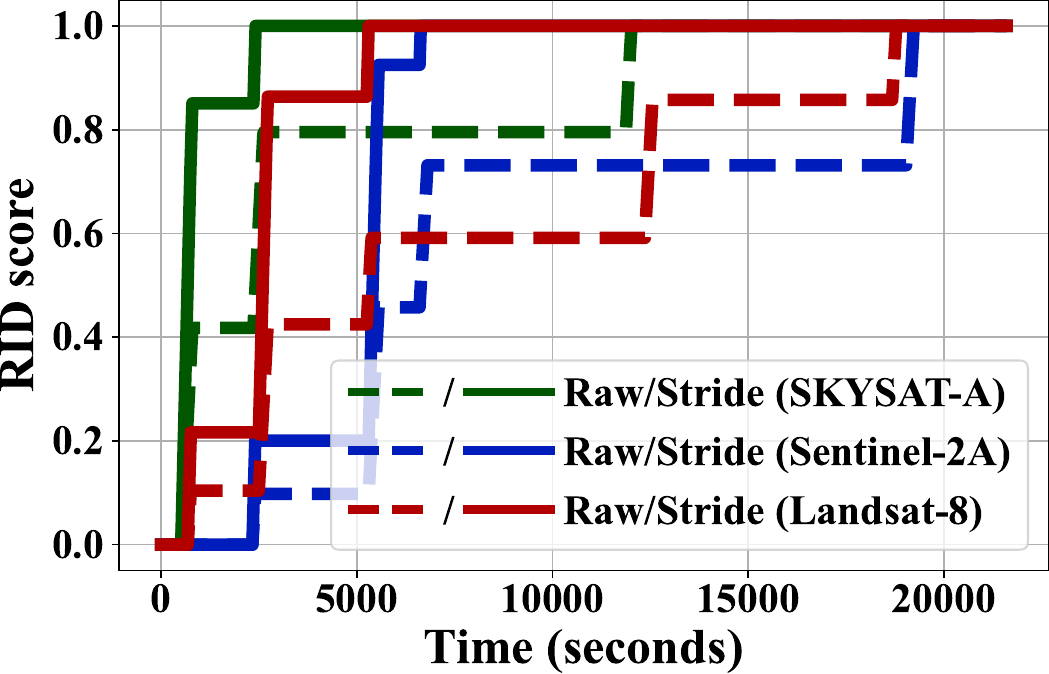}  \label{fig:Use_Case1_2}}
\vspace{-0.05in}
\caption{Use case 1: low-latency connectivity.}
\vspace{-0.1in}
\label{fig:Use_Case1}
\end{minipage}
\hfill
\begin{minipage}{0.49\linewidth}
\centering
    \subfloat[Transmission load over tiles.]{\includegraphics[width=0.5\textwidth]{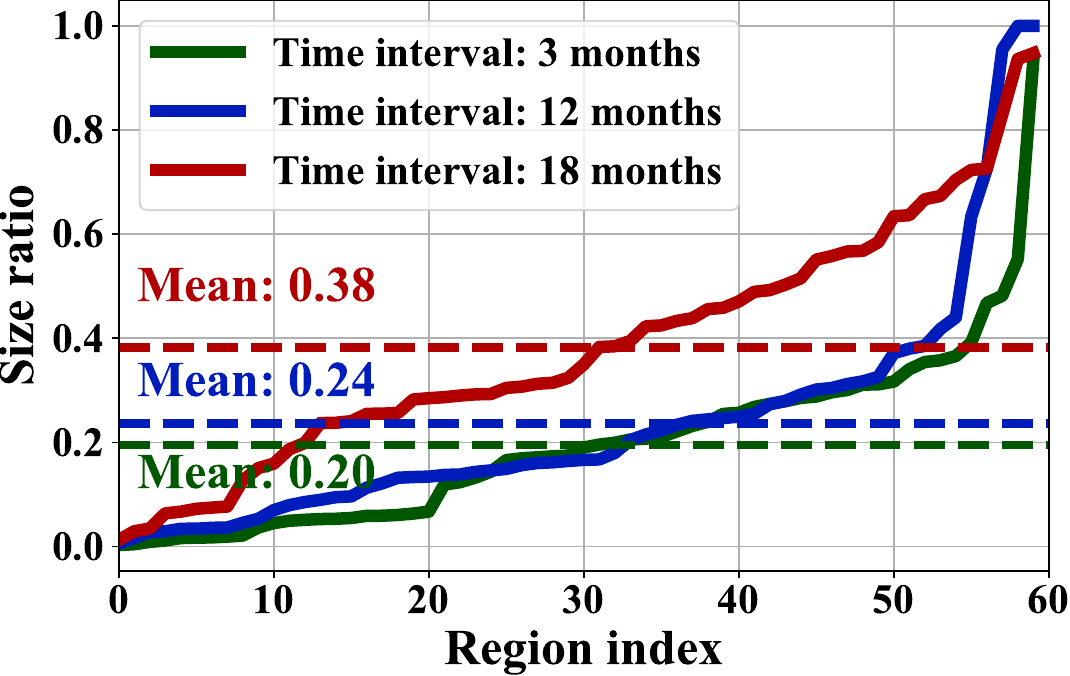}  \label{fig:Use_Case2_1}} 
    \subfloat[Revisiting imagery delivery.]{\includegraphics[width=0.5\textwidth]{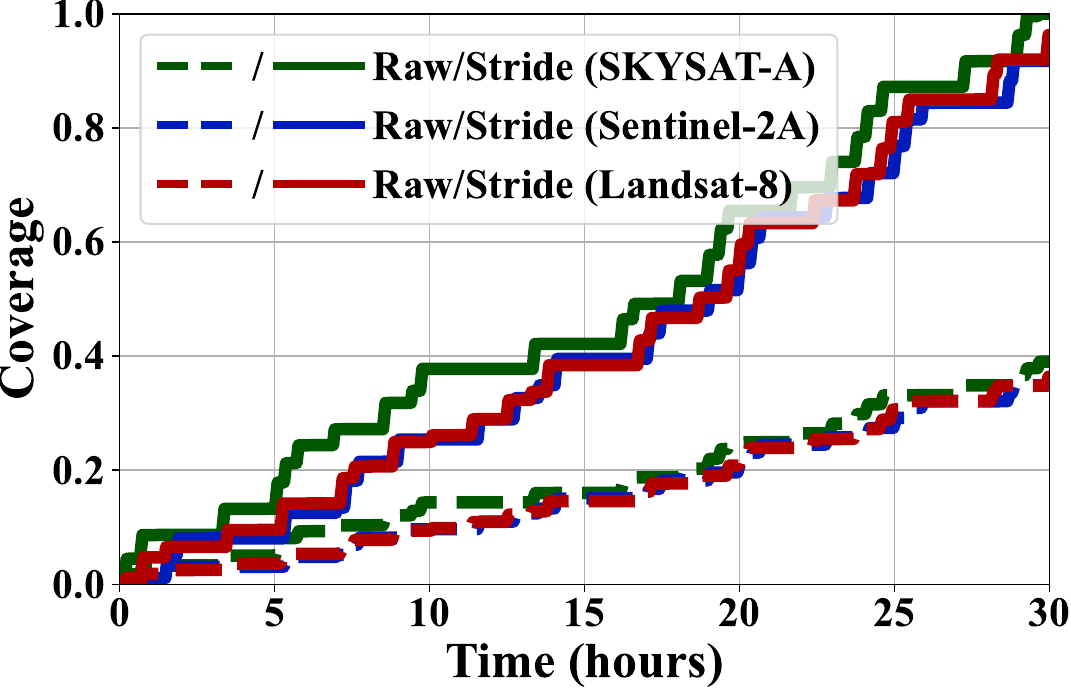}  \label{fig:Use_Case2_2}} 
\vspace{-0.05in}
\caption{Use case 2: wide-coverage mapping.}
\vspace{-0.1in}
\label{fig:Use_Case2}
\end{minipage}
\end{figure*}

\subsection{Quantitative Comparison} \label{sec:Quantitative Comparison}
We conduct quantitative comparisons against the aforementioned benchmarks on the \textit{DynamicEarthNet}~\cite{toker2022dynamicearthnet} dataset under part of random regions. 

\noindent
\textbf{Lossless settings.}
To ensure a fair comparison, we evaluate \System against JPEG-2000~\cite{taubman2002jpeg2000} and CCSDS 121.0-B-3~\cite{CCSDS121B3} under lossless mode (i.e., mean SSIM $=1.00$). 
As shown in~\cref{fig:Comparison1}, \System consistently outperforms all baselines, achieving size ratios of 0.51, 0.40, and 0.22 for CCSDS 121.0-B-3~\cite{CCSDS121B3}, JPEG-2000~\cite{taubman2002jpeg2000}, and \System, respectively. 
This corresponds to compression gains of 1.94$\times$, 2.51$\times$, and 4.62$\times$. 
Furthermore, \cref{fig:Comparison2} presents the distribution of size ratios, where \System demonstrates a clear concentration toward lower values, indicating superior compression efficiency.

\noindent
\textbf{Lossy settings.}
Likewise, we compare \System with Earth+~\cite{du2025earth+} under lossy settings. 
For fairness, both methods are evaluated using the same reference images and constrained to high reconstruction quality, with mean SSIM values consistently above 0.99 (approaching 1.00). 
As illustrated in~\cref{fig:Comparison3} and~\cref{fig:Comparison4}, \System significantly outperforms Earth+~\cite{du2025earth+}, achieving a performance gain of approximately 1.69$\times$ (i.e., 36.05$\times$ vs. 21.36$\times$), which can be attributed to a finer-grained operational design. 
Overall, these results highlight the superior capability and robustness of \System in practical scenarios.

\subsection{Use Cases} \label{sec:Use Cases}
\subsubsection{Case 1: Low-Latency Connectivity} \label{sec:Case 1: Low-Latency Connectivity}
We study low-latency connectivity on the \textit{SatSOT}~\cite{zhao2022satsot} dataset, where satellites continuously monitor an area and transmit video frames to ground stations, with \System using a fixed reference interval. 
\cref{fig:Use_Case1_1} shows that \System substantially reduces the transmission size ratio.
The curve exhibits a cyclic pattern over each reference interval: peaks correspond to reference-image transmission, while valleys occur shortly afterward when image changes are minimal; as time passes, both the changes and the transmission load increase with the varying spatial extent. 
\cref{fig:Use_Case1_2} shows that \System significantly reduces the revisiting imagery delivery time across satellite types, e.g., by 5.02$\times$ over \Raw on the SKYSAT-A satellite.
Notably, the gain is not proportional to the size reduction, attributed to the misalignment between the satellite trajectories and the distribution of ground stations.

\subsubsection{Case 2: Wide-Coverage Mapping}\label{sec:Case 2: Wide-Coverage Mapping}
We study wide-coverage mapping on the \textit{SpaceNet 7}~\cite{van2021spacenet} dataset, where satellites transmit bitemporal images of region tiles within a fixed time range, and coverage is defined as the ratio of transmitted images containing all ones.
\cref{fig:Use_Case2_1} shows that \System substantially reduces the transmission size ratio, which increases with the time interval as larger temporal gaps lead to greater spatial differences.
\cref{fig:Use_Case2_2} shows that \System significantly expands coverage across satellite types, achieving up to 2.56$\times$ improvement over \Raw on SKYSAT-A.

\section{Discussion} \label{sec:Discussion}
\noindent
\textbf{Storage analysis.}
\System requires additional onboard storage capacity to archive historical revisiting images. 
Theoretically, \System necessitates slightly more than a single-fold imagery storage capacity within each revisiting cycle to accommodate reference images. 
However, the data volume here refers to raw imagery before compression, which is manageable given satellite onboard storage can reach several terabytes~\cite{duggan2025advancing}. 


\noindent
\textbf{Fault analysis.}
Regarding fault problems~\cite{li2024satguard}, \System is designed to effectively prevent fault cases through the co-design of satellites and ground stations in terms of information sharing and decision checking, independent of the cloud indicator or reference selector. 
In cases of any wrong cloud indicator or lack of valid reference, \System directly transmits the image to bypass any potential faults.

\section{Related Work} \label{sec:Related Work}
\noindent
\textbf{In-orbit edge computing.}
In-orbit edge computing has emerged as a promising paradigm in recent years~\cite{tao2023transmitting, tao2024known, xing2024deciphering, liu2024orbit}, particularly for addressing the downlink bandwidth bottleneck in remote sensing systems. 
Kodan~\cite{denby2023kodan} utilizes a one-time transformation step and model specialization step for intelligent imagery filtering, which violates data completeness. 
DeepSpace~\cite{sun2025deepspace} downsamples satellite imagery onboard via simple resizing and reconstructs it on the ground using a trained wavelet diffusion model, which violates data reliability. 
Earth+~\cite{du2025earth+} relies on uplinking historical reference images from the ground to compress the captured images onboard, which lacks deployability. 
Uniquely, \System leverages the revisiting properties to resolve the downlink bandwidth bottlenecks. 

%


\noindent
\textbf{Satellite communication and networking.}
Satellite networks are becoming an increasingly significant topic, attracting considerable attention across various domains, such as communication~\cite{sun2025spacesched, pan2023pmsat, singh2024spectrumize, singh2021community}, networking stacks~\cite{shenoy2024cosmac, cao2023satcp, yang2016towards, li2024stable, liu2024dark, wu2025sate}, and services~\cite{liu2024democratizing, li2024dual, ren2024sateriot, liu2015co, li2024plug, dong2024gpsense}. 
In parallel, \System architects a revisiting-centric communication system from the perspective of celestial mechanics.

\section{Conclusion} \label{sec:Conclusion}
In this paper, we propose the design, implementation, and evaluation of \System, the first revisiting-aware in-orbit edge computing framework for Earth observation. 
\System introduces a cloud indicator, a reference selector, and a change detector to operate under conditions of meteorological contamination, orbit deviation, and intrinsic inter-band complexities \& data perturbations. 
The evaluation results show that \System improves the RID score by up to 4.55$\times$, decreases the connectivity latency by 5.02$\times$, and enlarges the mapping coverage by 2.56$\times$. 
Moreover, \System yields state-of-the-art performance with existing solutions, which presents a transformative impact on modern satellite systems. 


\bibliographystyle{IEEEtran}
\bibliography{main}

@misc{web:Statista,
  author={Statista},
  title = {Number of active satellites from 1957 to 2022},
  howpublished = {\url{https://www.statista.com/statistics/897719/number-of-active-satellites-by-year/}},
  year={2023}
}

@article{russell1964kepler,
  title={Kepler's laws of planetary motion: 1609--1666},
  author={Russell, John L},
  journal={The British journal for the history of science},
  year={1964}
}

@article{nadoushan2015repeat,
  title={Repeat ground track orbit design with desired revisit time and optimal tilt},
  author={Nadoushan, Mahdi Jafari and Assadian, Nima},
  journal={Aerospace Science and technology},
  year={2015}
}

@article{justice2002overview,
  title={An overview of MODIS Land data processing and product status},
  author={Justice, Christopher O and Townshend, John RG and Vermote, Eric F and Masuoka, Edward and Wolfe, Robert E and Saleous, Nazmi and Roy, David P and Morisette, Jeffrey T},
  journal={RSE},
  year={2002}
}

@article{li2022cloud,
  title={Cloud and cloud shadow detection for optical satellite imagery: Features, algorithms, validation, and prospects},
  author={Li, Zhiwei and Shen, Huanfeng and Weng, Qihao and Zhang, Yuzhuo and Dou, Peng and Zhang, Liangpei},
  journal={ISPRS Journal of P\&RS},
  year={2022}
}

@article{zhu2015improvement,
  title={Improvement and expansion of the Fmask algorithm: Cloud, cloud shadow, and snow detection for Landsats 4--7, 8, and Sentinel 2 images},
  author={Zhu, Zhe and Wang, Shixiong and Woodcock, Curtis E},
  journal={RSE},
  year={2015}
}

@article{wang2006cloud,
  title={Cloud masking for ocean color data processing in the coastal regions},
  author={Wang, Menghua and Shi, Wei},
  journal={IEEE TGRS},
  year={2006}
}

@article{jeppesen2019cloud,
  title={A cloud detection algorithm for satellite imagery based on deep learning},
  author={Jeppesen, Jacob H{\o}xbroe and Jacobsen, Rune Hylsberg and Inceoglu, Fadil and Toftegaard, Thomas Skj{\o}deberg},
  journal={RSE},
  year={2019}
}

@article{shao2019cloud,
  title={Cloud detection in remote sensing images based on multiscale features-convolutional neural network},
  author={Shao, Zhenfeng and Pan, Yin and Diao, Chunyuan and Cai, Jiajun},
  journal={IEEE TGRS},
  year={2019}
}

@article{weinberg1986orbital,
  title={Orbital decay of satellite galaxies in spherical systems},
  author={Weinberg, MARTIN D},
  journal={Astrophysical Journal},
  year={1986}
}

@article{weiss2018station,
  title={Station keeping and momentum management of low-thrust satellites using MPC},
  author={Weiss, Avishai and Kalabi{\'c}, Uro{\v{s}} V and Di Cairano, Stefano},
  journal={Aerospace Science and Technology},
  year={2018}
}

@article{cheng2024change,
  title={Change detection methods for remote sensing in the last decade: A comprehensive review},
  author={Cheng, Guangliang and Huang, Yunmeng and Li, Xiangtai and Lyu, Shuchang and Xu, Zhaoyang and Zhao, Hongbo and Zhao, Qi and Xiang, Shiming},
  journal={Remote Sensing},
  year={2024}
}

@article{treblow2024responsive,
  title={Responsive Maneuver Planning for Sun-Synchronous Repeating Ground Track Orbits},
  author={Treblow, Shalom and McGrath, Ciara N},
  journal={Journal of Spacecraft and Rockets},
  year={2024}
}

@inproceedings{bay2006surf,
  title={Surf: Speeded up robust features},
  author={Bay, Herbert and Tuytelaars, Tinne and Van Gool, Luc},
  booktitle={Proceedings of ECCV},
  year={2006}
}

@article{thacker1989role,
  title={The role of the Hessian matrix in fitting models to measurements},
  author={Thacker, William Carlisle},
  journal={Journal of Geophysical Research: Oceans},
  year={1989}
}

@article{stankovic2003haar,
  title={The Haar wavelet transform: its status and achievements},
  author={Stankovi{\'c}, Radomir S and Falkowski, Bogdan J},
  journal={Computers \& Electrical Engineering},
  year={2003}
}

@inproceedings{greathouse2014efficient,
  title={Efficient sparse matrix-vector multiplication on GPUs using the CSR storage format},
  author={Greathouse, Joseph L and Daga, Mayank},
  booktitle={Proceedings of ACM SC},
  year={2014}
}

@article{van2021spacenet,
  title={The spacenet multi-temporal urban development challenge},
  author={Van Etten, Adam and Hogan, Daniel},
  journal={arXiv preprint arXiv:2102.11958},
  year={2021}
}

@article{bai2021lscidmr,
  title={LSCIDMR: Large-scale satellite cloud image database for meteorological research},
  author={Bai, Cong and Zhang, Minjing and Zhang, Jinglin and Zheng, Jianwei and Chen, Shengyong},
  journal={IEEE Transactions on Cybernetics},
  year={2021}
}

@inproceedings{toker2022dynamicearthnet,
  title={Dynamicearthnet: Daily multi-spectral satellite dataset for semantic change segmentation},
  author={Toker, Aysim and Kondmann, Lukas and Weber, Mark and Eisenberger, Marvin and Camero, Andr{\'e}s and Hu, Jingliang and Hoderlein, Ariadna Pregel and {\c{S}}enaras, {\c{C}}a{\u{g}}lar and Davis, Timothy and Cremers, Daniel and others},
  booktitle={Proceedings of CVPR},
  year={2022}
}

@article{ma2023mtgl40,
  title={MTGL40-5: A Multi-Temporal Dataset for Remote Sensing Image Geo-Localization},
  author={Ma, Jingjing and Pei, Shiji and Yang, Yuqun and Tang, Xu and Zhang, Xiangrong},
  journal={Remote Sensing},
  year={2023}
}

@article{zhao2022satsot,
  title={SatSOT: A benchmark dataset for satellite video single object tracking},
  author={Zhao, Manqi and Li, Shengyang and Xuan, Shiyu and Kou, Longxuan and Gong, Shuai and Zhou, Zhuang},
  journal={IEEE TGRS},
  year={2022}
}

@article{taubman2002jpeg2000,
  title={JPEG2000: Standard for interactive imaging},
  author={Taubman, David S and Marcellin, Michael W},
  journal={Proceedings of the IEEE},
  year={2002}
}

@misc{CCSDS121B3,
  title        = {{Lossless Data Compression}},
  author       = {{Consultative Committee for Space Data Systems}},
  year         = {2020},
  howpublished = {\url{https://ccsds.org/Pubs/121x0b3.pdf}},
}

@article{duggan2025advancing,
  title={Advancing Earth observation: a survey on AI-powered image processing in satellites},
  author={Duggan, Aidan and Andrade, Bruno and Afli, Haithem},
  journal={European Journal of Remote Sensing},
  year={2025}
}

@inproceedings{denby2020orbital,
  title={Orbital edge computing: Nanosatellite constellations as a new class of computer system},
  author={Denby, Bradley and Lucia, Brandon},
  booktitle={Proceedings of ACM ASPLOS},
  year={2020}
}

@inproceedings{denby2023kodan,
  title={Kodan: Addressing the computational bottleneck in space},
  author={Denby, Bradley and Chintalapudi, Krishna and Chandra, Ranveer and Lucia, Brandon and Noghabi, Shadi},
  booktitle={Proceedings of ACM ASPLOS},
  year={2023}
}

@inproceedings{tao2024known,
  title={Known Knowns and Unknowns: Near-realtime Earth Observation Via Query Bifurcation in Serval},
  author={Tao, Bill and Chabra, Om and Janveja, Ishani and Gupta, Indranil and Vasisht, Deepak},
  booktitle={Proceedings of USENIX NSDI},
  year={2024}
}

@inproceedings{liu2024orbit,
  title={In-Orbit Processing or Not? Sunlight-Aware Task Scheduling for Energy-Efficient Space Edge Computing Networks},
  author={Liu, Weisen and Lai, Zeqi and Wu, Qian and Li, Hewu and Zhang, Qi and Li, Zonglun and Li, Yuanjie and Liu, Jun},
  booktitle={Proceedings of IEEE INFOCOM},
  year={2024}
}

@inproceedings{du2025earth+,
  title={Earth+: On-board satellite imagery compression leveraging historical earth observations},
  author={Du, Kuntai and Cheng, Yihua and Olsen, Peder and Noghabi, Shadi and Jiang, Junchen},
  booktitle={Proceedings of ACM ASPLOS},
  year={2025}
}

@inproceedings{sun2025deepspace,
  title={DeepSpace: Super Resolution Powered Efficient and Reliable Satellite Image Data Acquistion},
  author={Sun, Chuanhao and Zhang, Yu and Tao, Bill and Vasisht, Deepak and Marina, Mahesh},
  booktitle={Proceedings of ACM SIGCOMM},
  year={2025}
}

@inproceedings{sun2025spacesched,
  title={SpaceSched: A Constellation-Wide Scheduling System for Resolving Ground Track Congestion in Remote Sensing},
  author={Sun, Zehua and Ni, Tao and Hu, Pengfei and Gu, Tao and Xu, Weitao},
  booktitle={Proceedings of ACM MobiCom},
  year={2025}
}

@inproceedings{tao2023transmitting,
  title={Transmitting, fast and slow: Scheduling satellite traffic through space and time},
  author={Tao, Bill and Masood, Maleeha and Gupta, Indranil and Vasisht, Deepak},
  booktitle={Proceedings of ACM MobiCom},
  year={2023}
}

@inproceedings{pan2023pmsat,
  title={PMSat: Optimizing Passive Metasurface for Low Earth Orbit Satellite Communication},
  author={Pan, Hao and Qiu, Lili and Ouyang, Bei and Zheng, Shicheng and Zhang, Yongzhao and Chen, Yi-Chao and Xue, Guangtao},
  booktitle={Proceedings of ACM MobiCom},
  year={2023}
}

@inproceedings{singh2021community,
  title={A community-driven approach to democratize access to satellite ground stations},
  author={Singh, Vaibhav and Prabhakara, Akarsh and Zhang, Diana and Ya{\u{g}}an, Osman and Kumar, Swarun},
  booktitle={Proceedings of ACM MobiCom},
  year={2021}
}

@inproceedings{singh2024spectrumize,
  title={Spectrumize: Spectrum-efficient Satellite Networks for the Internet of Things},
  author={Singh, Vaibhav and Chakraborty, Tusher and Jog, Suraj and Chabra, Om and Vasisht, Deepak and Chandra, Ranveer},
  booktitle={Proceedings of USENIX NSDI},
  year={2024}
}

@inproceedings{wu2025sate,
  title={SaTE: Low-Latency Traffic Engineering for Satellite Networks},
  author={Wu, Hao and Han, Yizhan and Rajpal, Mohit and Zhang, Qizhen and Wang, Jingxian},
  booktitle={Proceedings of ACM SIGCOMM},
  year={2025}
}

@inproceedings{shenoy2024cosmac,
  title={CosMAC: Constellation-Aware Medium Access and Scheduling for IoT Satellites},
  author={Shenoy, Jayanth and Chabra, Om and Chakraborty, Tusher and Jog, Suraj and Vasisht, Deepak and Chandra, Ranveer},
  booktitle={Proceedings of ACM MobiCom},
  year={2024}
}

@inproceedings{li2024stable,
  title={Stable Hierarchical Routing for Operational LEO Networks},
  author={Li, Yuanjie and Liu, Lixin and Li, Hewu and Liu, Wei and Chen, Yimei and Zhao, Wei and Wu, Jianping and Wu, Qian and Liu, Jun and Lai, Zeqi},
  booktitle={Proceedings of ACM MobiCom},
  year={2024}
}

@inproceedings{cao2023satcp,
  title={SaTCP: Link-Layer Informed TCP Adaptation for Highly Dynamic LEO Satellite Networks},
  author={Cao, Xuyang and Zhang, Xinyu},
  booktitle={Proceedings of IEEE INFOCOM},
  year={2023}
}

@inproceedings{liu2024democratizing,
  title={Democratizing $\{$Direct-to-Cell$\}$ Low Earth Orbit Satellite Networks},
  author={Liu, Lixin and Li, Yuanjie and Li, Hewu and Yang, Jiabo and Liu, Wei and Lan, Jingyi and Wang, Yufeng and Li, Jiarui and Wu, Jianping and Wu, Qian and others},
  booktitle={Proceedings of USENIX NSDI},
  year={2024}
}

@inproceedings{li2024satguard,
  title={Satguard: Concealing endless and bursty packet losses in leo satellite networks for delay-sensitive web applications},
  author={Li, Jihao and Li, Hewu and Lai, Zeqi and Wu, Qian and Liu, Yijie and Zhang, Qi and Li, Yuanjie and Liu, Jun},
  booktitle={Proceedings of ACM WWW},
  year={2024}
}

@article{yang2016towards,
  title={Towards energy-efficient routing in satellite networks},
  author={Yang, Yuan and Xu, Mingwei and Wang, Dan and Wang, Yu},
  journal={IEEE JSAC},
  year={2016}
}

@article{li2024dual,
  title={Dual Network Computation Offloading Based on DRL for Satellite-Terrestrial Integrated Networks},
  author={Li, Dongbo and Sun, Yuchen and Peng, Jielun and Cheng, Siyao and Yin, Zhisheng and Cheng, Nan and Liu, Jie and Li, Zhijun and Xu, Chenren},
  journal={IEEE TMC},
  year={2024}
}

@inproceedings{ren2024sateriot,
  title={SateRIoT: High-performance Ground-Space Networking for Rural IoT},
  author={Ren, Yidong and Gamage, Amalinda and Liu, Li and Li, Mo and Chen, Shigang and Dong, Younsuk and Cao, Zhichao},
  booktitle={Proceedings of ACM MobiCom},
  year={2024}
}

@article{liu2015co,
  title={CO-GPS: Energy efficient GPS sensing with cloud offloading},
  author={Liu, Jie and Priyantha, Bodhi and Hart, Ted and Jin, Yuzhe and Lee, Woosuk and Raghunathan, Vijay and Ramos, Heitor S and Wang, Qiang},
  journal={IEEE TMC},
  year={2015}
}

@inproceedings{li2024plug,
  title={Plug-and-play Indoor GPS Positioning System with the Assistance of Optically Transparent Metasurfaces},
  author={Li, Ruinan and Zheng, Xiaolong and Liu, Liang and Ma, Huadong},
  booktitle={Proceedings of ACM MobiCom},
  year={2024}
}

@inproceedings{dong2024gpsense,
  title={GPSense: Passive Sensing with Pervasive GPS Signals},
  author={Dong, Huixin and Cui, Minhao and Wang, Ning and Qiu, Lili and Xiong, Jie and Wang, Wei},
  booktitle={Proceedings of ACM MobiCom},
  year={2024}
}

@inproceedings{xing2024deciphering,
  title={Deciphering the enigma of satellite computing with cots devices: Measurement and analysis},
  author={Xing, Ruolin and Xu, Mengwei and Zhou, Ao and Li, Qing and Zhang, Yiran and Qian, Feng and Wang, Shangguang},
  booktitle={Proceedings of ACM MobiCom},
  year={2024}
}

@inproceedings{liu2024dark,
  title={The dark side of scale: Insecurity of direct-to-cell satellite mega-constellations},
  author={Liu, Wei and Li, Yuanjie and Li, Hewu and Chen, Yimei and Wang, Yufeng and Lan, Jingyi and Wu, Jianping and Wu, Qian and Liu, Jun and Lai, Zeqi},
  booktitle={Proceedings of IEEE S\&P},
  year={2024}
}

@article{hsu2024real,
  title={Real-Time Compressed Sensing for Joint Hyperspectral Image Transmission and Restoration for CubeSat},
  author={Hsu, Chih-Chung and Jian, Chih-Yu and Tu, Eng-Shen and Lee, Chia-Ming and Chen, Guan-Lin},
  journal={IEEE TGRS},
  year={2024}
}

\end{document}